\documentclass[11pt,a4paper]{article}
\pdfoutput=1
\usepackage{jheppub}

\usepackage{tabularx,afterpage,xifthen}
\usepackage{amsmath,amsfonts,amssymb,mathtools}
\usepackage{slashed}
\usepackage{comment}
\usepackage{tikz}
\usepackage{caption}
\usepackage{subcaption}
\usepackage{graphicx, xcolor}
\newcommand{\bea}{\begin{eqnarray}}  
\newcommand{\eea}{\end{eqnarray}}

\newcommand{\nc}{\newcommand}
\nc{\beq}{\begin{equation}}
\nc{\eeq}{\end{equation}}
\nc{\barray}{\begin{eqnarray}}
\nc{\earray}{\end{eqnarray}}
\nc{\barrayn}{\begin{eqnarray*}}
\nc{\earrayn}{\end{eqnarray*}}
\nc{\bcenter}{\begin{center}}
\nc{\ecenter}{\end{center}}
\nc{\mc}{\mathcal}
\nc{\er}[1]{(\ref{eq:#1})}
\nc{\onehalf}{\frac{1}{2}} 
\nc{\partialbar}{\bar{\partial}}
\nc{\psit}{\widetilde{\psi}}
\nc{\hc}{\mbox{H.c.}}
\nc{\ev}{\;\mathrm{eV}}
\nc{\mev}{\;\mathrm{MeV}}

\nc{\gev}{\;\mathrm{GeV}}
\nc{\kev}{\;\mathrm{keV}}
\nc{\tev}{\;\mathrm{TeV}}

\newcommand{\fref}[1]{Fig.~\ref{#1}}
\newcommand{\eref}[1]{Eq.~(\ref{#1})}

\newcommand{\sref}[1]{Section~\ref{#1}}

\newcommand{\cref}[1]{Chapter~\ref{#1}}

\newcommand{\GK}[1]{{\bf\color{green}[GK]}} 

\title{
Unsupervised Hadronic SUEP at the LHC
}
\author{Jared Barron,$^{a}$}
\emailAdd{jared.barron@mail.utoronto.ca}

\author{David Curtin,$^{a}$}
\emailAdd{aspourda@physics.utoronto.ca}

\author{Gregor Kasieczka,$^{b}$}
\emailAdd{gregor.kasieczka@cern.ch }

\author{Tilman Plehn,$^{c}$}
\emailAdd{plehn@uni-heidelberg.de}

\author{and Aris Spourdalakis$^{a}$}
\emailAdd{aspourda@physics.utoronto.ca}

\affiliation{$^{a}$Department of Physics, University of Toronto, Toronto, Ontario, Canada M5S 1A7}
\affiliation{$^{b}$Institut f\"ur Experimentalphysik, Universit\"at Hamburg, Germany}
\affiliation{$^{c}$Institut f\"ur Theoretische Physik, Universit\"at Heidelberg, Germany}

\abstract{
Confining dark sectors with pseudo-conformal dynamics produce SUEPs, or Soft Unclustered Energy Patterns,
at colliders: isotropic dark hadrons with soft and democratic energies.
We target the experimental nightmare scenario, SUEPs in exotic Higgs decays, where all dark hadrons decay promptly to SM hadrons.
First, we identify three promising observables: the charged particle multiplicity, the event ring isotropy, and the matrix of geometric distances between charged tracks.
Their patterns can be exploited through a cut-and-count search, supervised machine learning, or an unsupervised autoencoder. 
We find that the HL-LHC will probe exotic Higgs branching ratios at the per-cent level, even without a detailed knowledge of the signal features.
Our techniques can be applied to other SUEP searches, especially the unsupervised strategy, which is independent of 
overly specific model assumptions and the corresponding precision simulations.
}

\begin{document}

\maketitle

%%%%%%%%%%%%%%%%%%%%%%%%%%%%%%
%%%%%%%%%%%%%%%%%%%%%%%%%%%%%%
%%%%%%%%%%%%%%%%%%%%%%%%%%%%%%
%%%%%%%%%%%%%%%%%%%%%%%%%%%%%%
%%%%%%%%%%%%%%%%%%%%%%%%%%%%%%
%%%%%%%%%%%%%%%%%%%%%%%%%%%%%%

\section{Introduction}

Hidden sectors are one of the most interesting and generic paradigms for physics beyond the Standard Model (BSM). 
These kinds of new particles and forces are not only plausible from a bottom-up point of view, but also arise in many top-down BSM theories~\cite{Strassler:2006im, Strassler:2008fv,Chacko:2005pe,Schabinger:2005ei,Patt:2006fw,Espinosa:2007qk,March-Russell:2008lng,Alimena:2019zri,Curtin:2021alk,Holdom:1985ag,Abel:2008ai,Batell:2009yf,Jaeckel:2010ni,Foot:2014mia,Feldman:2007wj,Pospelov:2007mp,Dudas:2013sia,An:2012va,Kribs:2018ilo, Knapen:2021eip, Cvetic:2002qa,Hur:2007uz,Bai:2013xga,Grossman:2010iq}.
In most scenarios of interest, hidden sector particles couple to SM fields via feeble interactions or heavy messengers, and the nature of these portal couplings determines their collider phenomenology.
A case of special interest are hidden valleys~\cite{Strassler:2006im}, referring to  hidden sectors with a confining gauge group which gives rise to rich infrared (IR) dynamics from very simple ultraviolet (UV) theory structures. 
The production of dark quarks at the LHC leads to a dark shower and high-multiplicity production of dark hadrons, in analogy to QCD jets. Depending on the portal by which the dark hadrons are produced and decay, these dark showers produce a wide variety of LHC signatures, which have been the subject of intense theoretical and experimental study in the last decade~\cite{Han:2007ae,Strassler:2008fv,Buschmann:2015awa,Arkani-Hamed:2008kxc,Ellis:2012zp,Toro:2012sv,ATLAS:2018dfo,Alimena:2019zri,Cohen:2015toa,Burdman:2018ehe,Schwaller:2015gea, Cohen:2017pzm, Cohen:2020afv}.

We consider one of the most challenging varieties of dark showers, Soft Unclustered Energy Patterns (SUEPs). If a hidden valley possesses a large gauge coupling that is pseudo-conformal above its confinement scale, then large-angle emission is unsuppressed for most of the parton shower evolution. This means the dark hadrons are not arranged in narrow QCD-like jets, but emitted approximately isotropically in the shower centre-of-mass frame~\cite{Strassler:2008bv,Knapen:2016hky}. This defines the SUEP final state as a high-multiplicity spherically-symmetric shower of hidden sector states. 
While existing searches can be sensitive to SUEPs produced at high energy scales, or with dark hadrons that decay to sufficiently conspicuous final states like leptons or Long-Lived Particles (LLPs)~\cite{Knapen:2016hky, Alimena:2019zri}, no dedicated SUEP searches exist to date. Furthermore, SUEPs \emph{without} conspicuous final states are not captured by any existing search and represent an unusually cruel signal, since their soft, isotropic distributions can mimic the ubiquitous pile-up produced by simultaneous LHC collisions.

To ensure that all types of SUEP signals can be discovered at the LHC, we focus on a well-motivated SUEP nightmare scenario, where SUEP is produced in exotic Higgs decays and the dark hadrons decay promptly and exclusively to SM hadrons. 
The modest energy scale of exotic Higgs decays and the lack of conspicuous final states forces us to rely on the kinematics of the resulting SM hadrons to extract the signal from the overwhelming QCD background. This production mode also allows us to side-step the problem of how to trigger on SUEPs~\cite{Knapen:2016hky} by using leptons in associated $Vh$ production. The  analysis techniques we develop will not only allow for the detection of this SUEP nightmare scenario, but should also increase the LHC experiments' sensitivity to all other SUEP possibilities.

An acute obstacle for SUEP searches is the lack of rigorous predictions and simulations for the strongly coupled pseudo-conformal dark sectors. Rather than hoping for conspicuous final states, we utilize the kinematics of the SM hadrons without relying on fine details of the signal beyond the robust SUEP characteristics of isotropic, soft, and democratic dark hadron energies. We therefore simulate SUEP production using a simple 
QCD fireball model of thermal dark hadron emission~\cite{Knapen:2021eip}, and find robust observables and event representations. An important observable for SUEP searches is the inter-particle distance matrix $\Delta R_{ij}$ between charged hadrons. This matrix encodes the essential geometric differences between QCD-like and SUEP-like hadron production. It forms the backbone of all our analysis strategies, together with known variables like event isotropy~\cite{Cesarotti:2020hwb} and the charged particle multiplicity.

To demonstrate the distinguishing power of our observables, as well as the drastic improvements from more sophisticated techniques, we examine three strategies for systematic SUEP searches at the HL-LHC. First, we simulate a simple cut-and-count analysis, which will turn out to allow for impressive sensitivities to $\mathrm{Br}(h \to \mathrm{SUEP}) \sim 1\%$ at the HL-LHC. We anticipate that a realistic analysis with data-driven background estimation will perform even better, since our study is limited by background simulation statistics. To improve on this, we utilize supervised as well as unsupervised machine learning (ML).
Unsupervised analysis concepts along the lines of autoencoder neural networks~\cite{Rumelhart1986} have the potential to transform LHC analyses~\cite{Asadi:2017qon,Metodiev:2017vrx,Andreassen:2018apy,Collins:2018epr,DeSimone:2018efk,Heimel:2018mkt,Farina:2018fyg,Roy:2019jae,Cheng:2020dal,Nachman:2020lpy,MdAli:2020yzb,Bortolato:2021zic,Dillon:2021nxw,Finke:2021sdf,Kasieczka:2021xcg,Aarrestad:2021oeb,Cerri:2018anq}, including 
searches for dark showers~\cite{Heimel:2018mkt}. We point out how unsupervised methods are especially appealing for difficult-to-simulate signals like SUEP since they only rely on the known QCD background for training, while yielding at least several times greater SUEP sensitivity than the cut-and-count analysis. 

Our investigation establishes that SUEP searches need not rely on conspicuous SM final states for excellent sensitivity at the HL-LHC. The unique dark hadron kinematics, which robustly follows from their origin in pseudo-conformal strongly coupled dynamics, allows for the SUEP final state to be distinguished from its overwhelming QCD background. Our techniques can be applied to all SUEP searches to dramatically enhance their sensitivity, regardless of energy scale or SM final state.

This paper is structured as follows.
In \sref{sec:SUEP}, we briefly review the SUEP theory and define the benchmark scenario for our study. 
Signal and background simulation is discussed in in \sref{sec:simulation}.
In \sref{sec:analysis} we define the relevant observables to distinguish SUEP from QCD and discuss supervised and unsupervised machine learning techniques.
In \sref{sec:results} we present our results, including projections for the $\mathrm{Br}(h\to \mathrm{SUEP})$ sensitivity at the HL-LHC, and we conclude in \sref{sec:Summary}.

%%%%%%%%%%%%%%%%%%%%%%%%%%%%%%
%%%%%%%%%%%%%%%%%%%%%%%%%%%%%%
%%%%%%%%%%%%%%%%%%%%%%%%%%%%%%
%%%%%%%%%%%%%%%%%%%%%%%%%%%%%%
%%%%%%%%%%%%%%%%%%%%%%%%%%%%%%
%%%%%%%%%%%%%%%%%%%%%%%%%%%%%%
\section{Theory of SUEPs}\label{sec:SUEP}

Hidden valley models are a large class of BSM theories in which the SM is extended by additional gauge groups under which SM particles are neutral. New particles that are charged under non-Abelian extensions can give rise to a wide range of interesting hidden sector dynamics~\cite{Strassler:2006im,Strassler:2008fv,Han:2007ae} and various challenging SM signatures.
These models appear in the context of many top-down constructions~\cite{Morrissey:2009tf}, including string theory, of course~\cite{Cvetic:2002qa}, and are compatible with various potential resolutions to the hierachy problem such as supersymmetry \cite{Arkani-Hamed:2005zuc}, little Higgs models, TeV extra dimensions and Randall-Sundrum scenarios \cite{Strassler:2006im,Arkani-Hamed:2001nha,Randall:1999ee,Randall:1999vf}. In Neutral Naturalness scenarios, hidden valleys actually solve the little hierarchy problem directly~\cite{Chacko:2005pe}. Models containing a hidden valley have also been studied in the context of dark matter~\cite{Hur:2007uz},  matter-antimatter asymmetry~\cite{Bai:2013xga} and the origin of neutrino masses~\cite{Grossman:2010iq}. 

In most hidden sector scenarios with a confining gauge force, the dark parton shower is qualitatively similar to QCD: the asymptotic freedom of the running gauge coupling enhances soft and collinear emission, resulting in the production of hidden sector states in collimated jets. If the hidden 't Hooft coupling is large ($\lambda\equiv g^2 N_c \gg 1$) and approximately constant over a significant energy range, the distribution of the produced dark hadron momenta will be much more democratic than the hierarchical jet-like behaviour we see in QCD. This is the SUEP class of signals, characterized by relatively soft, isotropic emission of dark hadrons \cite{Strassler:2008bv,Knapen:2016hky}. 
It is worth keeping in mind that strongly coupled hidden sector dynamics is not the only scenario that can lead to SUEPs. Similar final states  can be produced in many-step cascade decays in the hidden sector~\cite{Elor:2015tva}, or in theories with extra spatial dimensions (see e.g.~\cite{Cesarotti:2020uod,Costantino:2020msc,Park:2012fe}). 
We now describe the general features of production, evolution and decay that constitute the SUEP signal, and define a benchmark scenario for our study.

\subsection{Dark Hadron Production in Exotic Higgs Decays}
Production of SUEP can occur through various portals coupling SM particles to SM-singlet states charged under a dark gauge group. Higgs and vector boson portals are commonly studied examples. Alternatively, a new particle charged under both SM and the dark gauge group could be produced~\cite{Knapen:2021eip}. We assume that hidden sector states $\psi_D$ can be produced via the Higgs portal:
\begin{equation}
    \mathcal{O}_\text{production}\sim |H|^2\psi_D \bar\psi_D \; .
\end{equation} 
$\psi_D$ could represent a fermion or scalar dark quark charged under the hidden gauge group. In the fermion case, the above operator is actually dimension 5 with a coupling ${\sim}1/Lambda$ for some UV scale $\Lambda$.  This operator will give rise to the production of hidden states in exotic Higgs decays, as long as the dark hadrons are kinematically accessible. The dark quarks hadronize into a large multiplicity of dark hadrons. 

For a confining gauge theory, the evolution of the system from the hard scale $Q$ at which the first partons charged under the gauge group are generated to the IR confinement scale $\Lambda$ governs the hadron multiplicity generated during showering. $Q$ is of the order $m_h$ in our case. For QCD jets, this evolution can be reliably simulated as a parton shower, but for SUEPs different strategies are required.

The average hadron multiplicity is given by
\begin{equation}
    \langle n (Q) \rangle=\int_0^1 F(x,Q)dx
\end{equation}
where $F$ is the fragmentation function describing the final state distribution of momenta~\cite{Ellis:1996mzs}, and $x$ is the momentum fraction of a given splitting. Calculating its evolution from the scale $Q$ down to an IR scale $\Lambda$ involves resumming the divergent contributions from the anomalous dimensions, with the leading contribution obtained from just the first ($j=1$) Mellin transform of the anomalous dimension. If the theory is conformal between $Q$ and $\Lambda$, the running of the coupling can (by definition) be neglected. In the limit where the 't Hooft coupling is large $\lambda\gg(1-x)$ it was first shown in Ref.~\cite{Hatta:2008tn} that
\begin{equation}
    \langle n \rangle\sim \left(\frac{Q}{\Lambda}\right)^{2\gamma_T(1)}  \; ,
\end{equation}
where $\gamma_T(1)$ is the first Mellin-Moment of the time-like anomalous dimension of the fragmentation function. 
Further expanding to zeroth order in the coupling yields
\begin{equation}
    \langle n \rangle\sim \left(\frac{Q}{\Lambda}\right)^{1 +\mathcal{O}(1/\sqrt{\lambda})}
\end{equation}
Note that the small momentum fraction $x$ carried by each individual splitting follows from $\lambda\gg (1-x)$. In this strong regime branching is expected to yield emissions with $x\sim \Lambda/Q$ that are relatively isotropic in direction and democratic in momentum~\cite{Hatta:2008tn, Knapen:2016hky}. 
Thus, with a large enough scale separation $Q \gg \Lambda$, low-$x$, high-multiplicity final states are generated. 
Branching terminates after $N_\text{final} \sim \mathrm{log}{\langle n \rangle}$ splittings at $Q_{N_\text{final}}\sim Q/2^{N_\text{final}} \sim \Lambda$, at which point hadronization takes over.

Hadron production in QCD at high temperatures is a close analogue to our situation, and statistical models have consistently shown that hadron multiplicities follow a thermal distribution~\cite{Fermi:1950jd,Hagedorn:1965st,PhysRevD.1.1416,Blanchard:2004du,Hatta:2008tn,Becattini:2001fg,Cleymans:2012wm,Becattini:2008tx,Becattini:2010sk,Becattini:2009ee,Ferroni:2011fh}. We use this picture as a toy model of dark hadron production in SUEP, modelling the distribution of dark meson momenta as a relativistic Boltzmann distribution  
\begin{equation}\label{eq:thermal_distro}
    \frac{dN}{d^{3}\bf{p}} \sim \exp\left(-\frac{\sqrt{{\bf{p}}^{2} + m_D^2}}{T_D} \right) \ ,
\end{equation}
where $m_{D}$ is the mass of the final dark states and $T_{D}$ acts as the Hagedorn temperature of the hidden confining gauge force, with $T_D \sim \Lambda$~\cite{Knapen:2016hky, Blanchard:2004du}. This temperature controls the kinetic energy of the dark hadron distribution.

\subsection{Dark Hadron Decay}

Generically the decay of the dark hadrons $\phi_D$ into the SM will occur through some effective coupling of the form 
\begin{equation}
    \mathcal{O}_\text{decay}=\phi_D \mathcal{O}_\text{SM}
\end{equation}
where $\mathcal{O}_\text{SM}$ contains fields charged under the SM gauge group.
The phenomenology of the SUEP signal is determined to a large extent by the dominant decay portal. 
For example, if the dark hadrons decay to massive dark photons that mix with the SM photon, then the SM final state contains both hadrons and leptons in roughly gauge-ordered proportions, though for various dark photon masses in the GeV-regime, hadrons can dominate~\cite{Knapen:2016hky}.
Alternatively, if the dark hadrons decay through the Higgs portal, the final state will contain hadrons and leptons in roughly Yukawa-ordered proportions.
For sufficiently small portal coupling, the decay length of the dark hadrons may also be macroscopic, resulting in LLP signatures~\cite{Alimena:2019zri}. 

It is also possible for the hidden sector states to decay purely hadronically, which is the most experimentally challenging case at the LHC. A very simple example is the gluon portal as described in \cite{Knapen:2021eip}:
\begin{equation}
        \mathcal{L}\supset -\frac{1}{2}m_a^2 a^2-\frac{\alpha_2}{8\pi}\frac{1}{f_a}aG_{\mu\nu}\tilde{G}^{\mu\nu}-iy_{\psi_D} a \psi_D \psi_D^* \ .
\end{equation}
Here, $a$ is a heavy elementary pseudo-scalar in the dark sector and $\psi_D$ is the dark quark. Dark hadrons, which are bound states of $\psi_D$, could then decay to SM hadrons via an effective operator $\phi_D G\tilde{G}$. Another example is the hadrophilic (or leptophobic) $Z'$ portal~\cite{Bernreuther:2019pfb}, where a new heavy gauge boson couples to SM quarks but not leptons, allowing dark hadrons to decay via an effective operator like $\phi_D q \bar q$.

\subsection{Prompt Hadronic Benchmark Scenario}

The production of hidden sector states via the Higgs portal generally and exotic Higgs decays in particular is one of the most motivated and plausible discovery scenarios for new physics~\cite{Curtin:2013fra}. 
It is therefore vital that our experimental search strategies cover all possibilities for a signal at the LHC.
This is especially urgent since for final states that are not covered by existing searches, branching fractions of ${\sim}10\%$ are easily allowed by current measurements of Higgs couplings and invisible decays~\cite{ATLAS:2018bnv,ATLAS:2019cid,ATLAS:2019nkf,CMS:2018yfx,Biekoetter:2018ypq}.

We therefore focus on the experimental worst-case scenario for SUEP produced in exotic Higgs decays: purely hadronic and prompt decays, with a particular interest in low dark hadron masses that make resonance searches~\cite{Pierce:2017taw, ATLAS:2020ahi} or applications of jet substructure techniques~\cite{Park:2017rfb} challenging. While the simplest gluon portal scenarios suggest that dark hadrons lighter than ${\sim}10 \gev$ have macroscopic decay lengths~\cite{Knapen:2021eip} (which could allow for the use of long-lived particle search techniques~\cite{Schwaller:2015gea,Alimena:2019zri}), other possibilities can easily realize prompt, purely hadronic decays over a much wider range of dark hadron masses. For example, the hadrophilic (leptophobic) vector portal~\cite{Bernreuther:2019pfb} with a hypothetical confining sector where the lightest dark hadron is a dark-rho-like vector $\rho_D$ would allow dark hadrons lighter than a GeV to decay promptly into SM hadrons.
In focusing on the prompt case we can develop techniques that allow SUEP production to be identified using only the geometrical and momentum distribution of its SM final states. These techniques will enhance our sensitivity for any kind of general SUEP, in addition to whatever other features of the final state, like displaced decays, leptons, or photons, can also be exploited.

\begin{figure}
    \centering
    \includegraphics[width=0.9\textwidth]{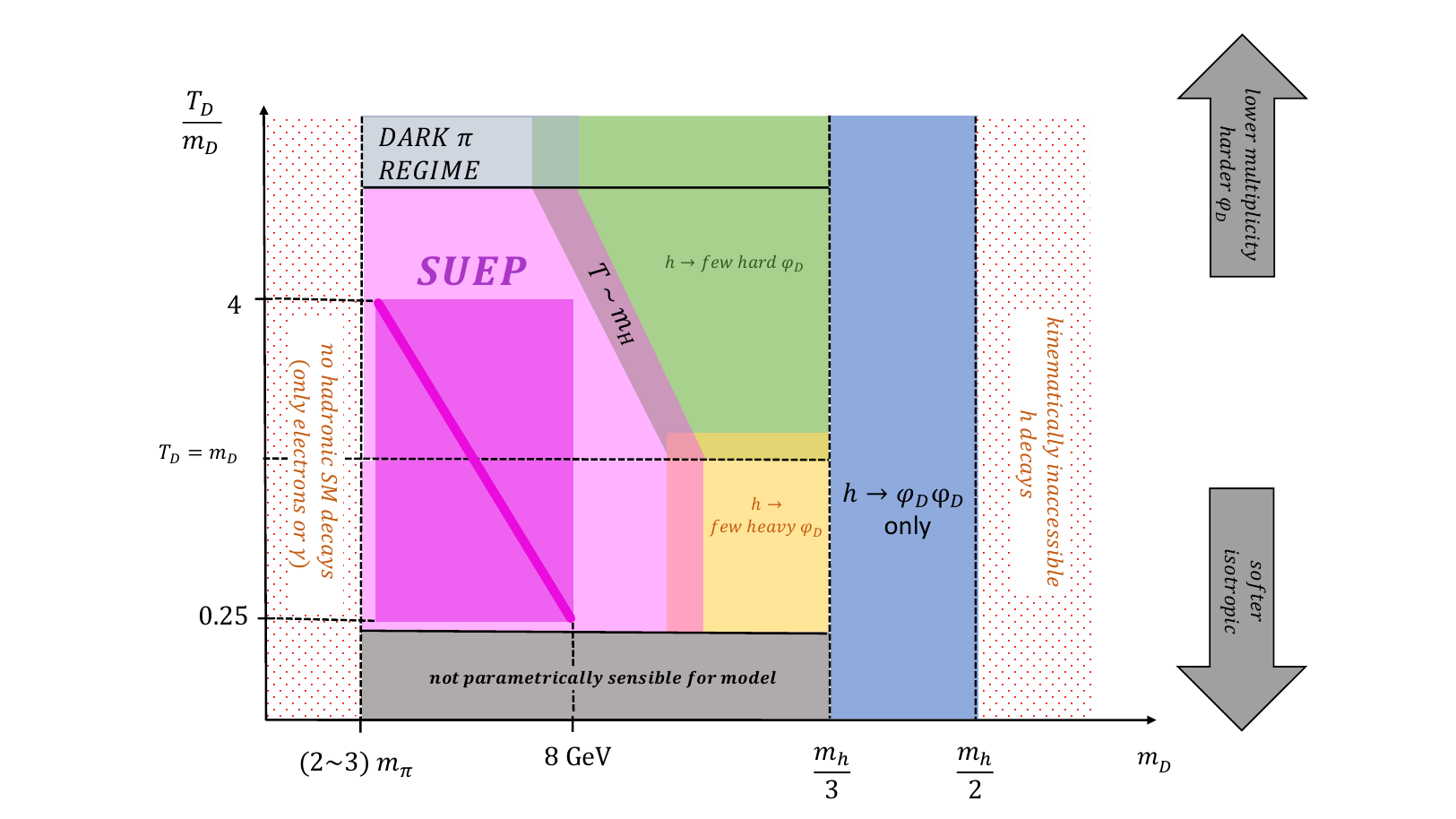}
    \caption{Cartoon of the our benchmark scenario for SUEP produced in Higgs decays with prompt decay of dark hadrons into SM hadrons. 
    The SUEP description applies in the purple area, for $T$ within a factor of a few of the dark hadron mass. 
    In the green region, dark hadron production is not thermal, but described by processes more analogous to chiral QCD. In the orange and green regions, searches for Higgs decays to a few resonances would be sensitive to this dark sector. 
    The parameter space region indicated with the darker purple rectangle is the focus of our analysis. Our cuts are optimized for the most archetypically SUEP-like final states, schematically indicated by the lower-left corner of this rectangle, demarcated with the diagonal line.
     }
    \label{fig:Cartoon}
\end{figure}

The parameter space of our benchmark scenario is shown schematically in Figure \ref{fig:Cartoon}.
The Higgs production portal sets the high scale for the event to $Q=125 \gev$. With this scale fixed, the distribution of final states is in principle determined by the dark hadron mass $m_{D}$ and the dark Hagedorn temperature $T_{D}$ of Eq. \eqref{eq:thermal_distro}, shown on the vertical and horizontal axes of 
Figure \ref{fig:Cartoon}.
In reality, there may be different dark hadrons with different spins, decays, and distributions, but this simplified description is sufficient for our purposes.
On the left and right, the relevant parameter space is kinematically bounded by the red-hatched areas. 
Dark hadron production in Higgs decays requires $m_D < m_h/2$. 
As we explain below, we focus on dark hadrons that decay to SM hadrons, which requires  $m_D > (2 - 3) m_\pi$, depending on the exact decay portal. 

Even within that mass range, hidden sectors with pseudo-conformal dynamics do not always manifest as SUEP signatures in exotic Higgs decays. 
If $m_D> {m_h}/{3}$ (blue area in Figure~\ref{fig:Cartoon}), then dark hadrons are only pair-produced in Higgs decays, making this scenario equivalent to standard exotic Higgs decays to pairs of various new particles (see e.g.~\cite{Curtin:2013fra}). 
If the dark hadron mass is fairly large, $m_D \sim m_h/(\mathrm{few})$ (yellow area), or the dark Hagedorn temperature is comparable to or above the Higgs mass $T \gtrsim m_h$ (green area), then exotic Higgs decay would produce only a small multiplicity of dark hadrons that are either fairly hard or fairly heavy. In both cases, dark hadron production is not thermal but is described by processes more akin to chiral QCD.
These regions are likely accessible through modified  searches for resonances in exotic Higgs decays~\cite{Pierce:2017taw, ATLAS:2020ahi}, and we do not focus on them here.
The gray area where $T/m \ll 1$ is not expected to be realized by any pseudo-conformal hidden sector, since the dark hadron mass and temperature are both related to the strong coupling scale $\Lambda$. 
On the other hand, $T/m \gg 1$ is possible in what we call the ``dark pion regime'', where the dark hadrons are pseudo-Goldstone bosons of an approximate symmetry, meaning their mass can be much smaller than $\Lambda$. We do not focus on this region, but it would be an interesting target for future investigations.

This leaves us with the actual SUEP regime for dark hadron production in exotic Higgs decays, indicated by the light purple area. In this investigation, we will focus on dark hadron masses below 8~GeV and $T/m$ in the reasonable range of $\sim$ 0.25 to 4. This target SUEP parameter space is marked out as the darker purple rectangle. Our cuts will be particularly optimized for the lower-left region of the rectangle, demarcated with the thick diagonal line. This is the region of low dark hadron mass and/or temperature, corresponding to the softest SM final states that are most difficult to search for using existing techniques.

%%%%%%%%%%%%%%%%%%%%%%%%%%%%%%
%%%%%%%%%%%%%%%%%%%%%%%%%%%%%%
%%%%%%%%%%%%%%%%%%%%%%%%%%%%%%
\section{Simulation}
\label{sec:simulation}

We briefly outline how we simulate our SUEP signal and the most important QCD backgrounds, where the latter is necessary to develop our analysis techniques, even though a realistic experimental analysis would rely on data-driven background estimation.

\subsection{Signal}\label{sec:signal_simulation}

We generate event samples for exotic Higgs decay into SUEP using the \texttt{SUEP\_Generator}  plugin~\cite{Knapen:2021eip} for \texttt{Pythia 8.243}~\cite{Sjostrand:2014zea}, which models the dark shower as a spherical distribution of dark pseudo-scalar mesons with momenta drawn from the relativistic Boltzmann distribution Eq. \eqref{eq:thermal_distro}. As in Ref.~\cite{Knapen:2016hky}, we make the simplifying assumption that there is only one flavor of dark meson produced in the exotic Higgs decay. The free parameters of the SUEP shower are the dark hadron mass $m_D$ and the effective temperature $T_D$, setting the energy scale at which dark hadrons are produced dominantly.

We simulate associated Higgs production at the 14~TeV HL-LHC, $pp \to V h, V \to \ell \ell/\ell \nu$,  in \texttt{Pythia 8}. The Higgs is decayed to the SUEP final state of dark mesons, which then decay directly to a $u \bar u$ quark pair that in turn undergoes SM hadronization. The exact choice of hadronic decay mode does not significantly affect our analysis, so we use this single channel as a stand-in for other purely hadronic portals. The events are then passed through the simplified detector simulation code \texttt{Delphes 3}~\cite{deFavereau:2013fsa} with CMS detector settings. The simulated detector-level objects output by Delphes are used for our analysis. Our SUEP search will only use charged-track information. Since charged tracks can be traced back to the primary vertex, they are very robust with respect to pile-up contamination. We can therefore neglect the effects of pile-up in the remainder of our study. 

In our event samples we cover $m_D$ from 400~MeV to 8~GeV, and $T_{D}/m_{D}$ from 0.25 to 4. The lower bound on the dark hadron mass ensures that the dark mesons are kinematically allowed to decay to two pions. The upper bound is chosen to show where our search loses sensitivity. The range of temperatures is chosen to satisfy $T_{D}\sim m_{D}$, which is the regime where the thermal picture of SUEP production is valid. The signal cross section is~\cite{Cepeda:2019klc} 
\begin{eqnarray}
\label{e.signalxsec}
\sigma(p p \to e\nu/\mu\nu + \ \mathrm{SUEP})
&=& (0.34 {\ \rm pb}) \cdot \mathrm{Br}(h\to \ \mathrm{SUEP})
\\
\nonumber 
\sigma(p p \to ee/\mu\mu + \ \mathrm{SUEP})
&=& (0.066 {\ \rm pb}) \cdot \mathrm{Br}(h\to \ \mathrm{SUEP})
\end{eqnarray}
for SUEP production in exotic Higgs decays in association with leptonic $W/Z$-bosons that decay into electrons or muons.
In total, we generate
$2.0\times 10^{5}$ $Zh$ and $Wh$ signal events, proportional to their respective cross section, 
 for each set of signal parameters $(m_D, T_D/m_D)$.

\subsection{Background}\label{sec:background}

The dominant background to our signal is production of one or two leptons in association with any number of QCD jets. It is highly challenging to model reliably, and in a realistic study, data-driven background estimation would be employed, see \sref{ss.datadriven}. However, for the purpose of developing our analysis techniques, we simulate QCD+leptons background samples using \texttt{MadGraph5\_aMC@NLO 2.6.6} and \texttt{Pythia 8.243}. 

Ideally, one should simulate fully matched multi-jet $+$ $\ell\ell/\ell\nu$ samples to capture the background distribution as closely as possible. However, due to the large statistics needed for our analysis, and the fact that such a simulation is anyway unlikely to be a perfect representation of the detailed hadronic distributions at the relevant high multiplicities and relatively low energy scale of the Higgs mass, this approach is not practical. 
Instead, we simulate $n j + \ell\ell/\ell\nu$, where $n = 2, 3, 4$ without jet matching and $p_T > 15$~GeV at generator level, to determine the effect of jet multiplicity at the hard event level on our analysis. We find that $n > 2$ leads to lower cross section while being \emph{more distinguishable} from the SUEP signal using the analysis techniques we develop here. Therefore, to be conservative, we simulate $2 j + \ell\ell/\ell\nu$ as our background samples for $Zh$ and $Wh$ production and decay into SUEP, respectively. In total, we use $10^{8}$ background events to represent the 
\begin{equation}
\label{e.bgxsec}
    \sigma(\mathrm{QCD} \  + \   \ell\ell/\ell\nu) \approx  3.7\times 10^{3} {\ \rm pb}
\end{equation}
lowest-order \texttt{MadGraph5} cross section for this background. While this is sufficient to develop our analysis techniques, the Monte Carlo background sample has $\sim 1/100$ the statistics of the full HL-LHC dataset. This is important for the interpretation of our results in \sref{sec:results}. 

%
%%%%%%%%%%%%%%%%%%%%%%%%%%%%%%
%%%%%%%%%%%%%%%%%%%%%%%%%%%%%%
%%%%%%%%%%%%%%%%%%%%%%%%%%%%%%
%%%%%%%%%%%%%%%%%%%%%%%%%%%%%%
%%%%%%%%%%%%%%%%%%%%%%%%%%%%%%
\section{Analysis}\label{sec:analysis}

The goal of this paper is to devise  strategies for extracting SUEP signals from a large background events without relying on the details of the simulated signal. We first describe our trigger assumptions and baseline cuts, and define a cut-based classifier to establish how sensitive such a simple approach can be. \sref{sec:supervised} introduces a supervised neural network classifier, to demonstrate both the advantages and limitations of the supervised approach for our physics problem. We then introduce our primary tool in \sref{sec:unsupervised} --- an unsupervised neural network that we employ as an anomaly detector for SUEP. 

\subsection{SUEP Observables}

We define our trigger pre-selection by requiring that all events have at least one charged electron or muon with $p_{T}\geq 40$ GeV, or two opposite sign charged leptons with $p_{T} \geq 30 (20)$~GeV. We also require that the scalar $p_T$-sum of hadronic charged tracks from \texttt{Delphes} is above $30$ GeV. Both signal and background have a trigger efficiency of ${\approx}40\%$, relative to the cross sections in \eref{e.signalxsec} and \eref{e.bgxsec}.

We focus on the most challenging region of SUEP parameter space, with either low dark hadron masses or low dark shower temperatures. This gives rise to the most archetypically SUEP-like final states with a high multiplicity of isotropically distributed, relatively soft SM hadrons. 
The three observables that best capture the characteristics of this signal are the charged particle multiplicity $N_\mathrm{charged}$, the event isotropy $\mathcal{I}$~\cite{Cesarotti:2020hwb}, and the interparticle distance. In all steps of the analysis that follow, we only use charged particle tracks with $p_{T}\geq 300$~MeV from the primary vertex, excluding the one or two hard leptons associated with the decaying gauge boson.
    
    The event isotropy observable $\mathcal{I}  \in (0, 1)$~\cite{Cesarotti:2020hwb} quantifies the energy mover's distance between a collider event and an idealized isotropic event with uniform energy distribution, so $\mathcal{I} = 0$ indicates a fully isotropic event. Originally, three different definitions of the event isotropy are laid out, utilizing different geometries --- spherical, cylindrical, or in a two-dimensional ring. We compute the \emph{ring isotropy} of the set of charged hadronic tracks of each event, since at a $pp$ collider we have no way to know the longitudinal boost of the Higgs that decays to SUEP. Since the SUEP is isotropic in the Higgs rest frame, we boost the hadronic charged track system of each event into its transverse rest-frame before computing the ring isotropy. To do this we assume that all hadronic charged tracks belong to particles with the pion mass, but this is sufficient to significantly separate signal and background events.
    
    The variable that we introduce for the specific purpose of studying SUEP events is the interparticle distance matrix $\Delta R_{ij}$ for charged hadron tracks in the lab frame.
    It captures the unique topology of SUEP events while being very suitable for machine-learning applications, since it is invariant under re-definitions of the azimuthal angle around the beam axis. It is also useful to define the mean 
    $\overline{\Delta R}$  of all matrix entries.

\begin{figure}[t]
     \includegraphics[width=0.335\textwidth]{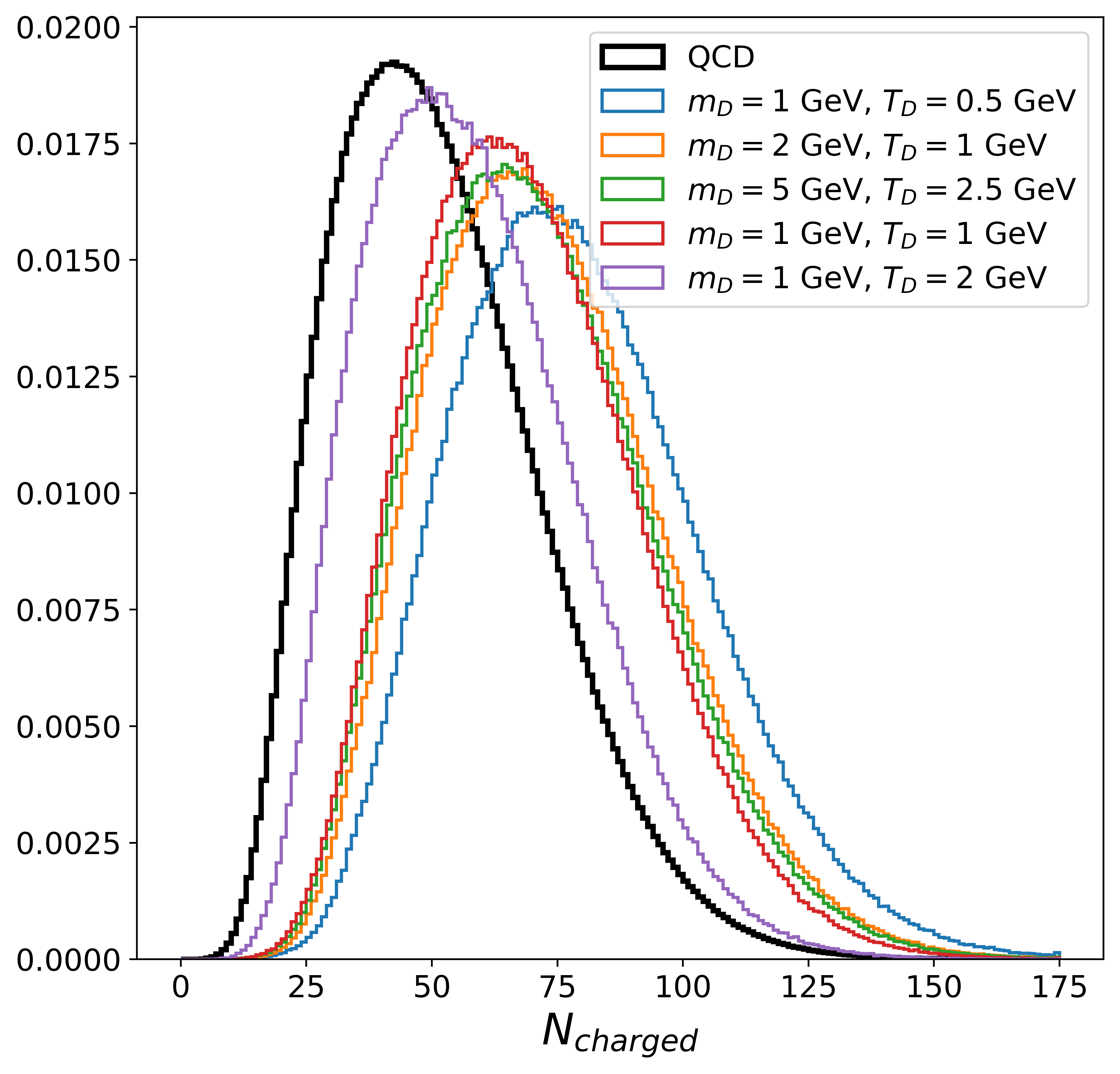}
     \includegraphics[width=0.31\textwidth]{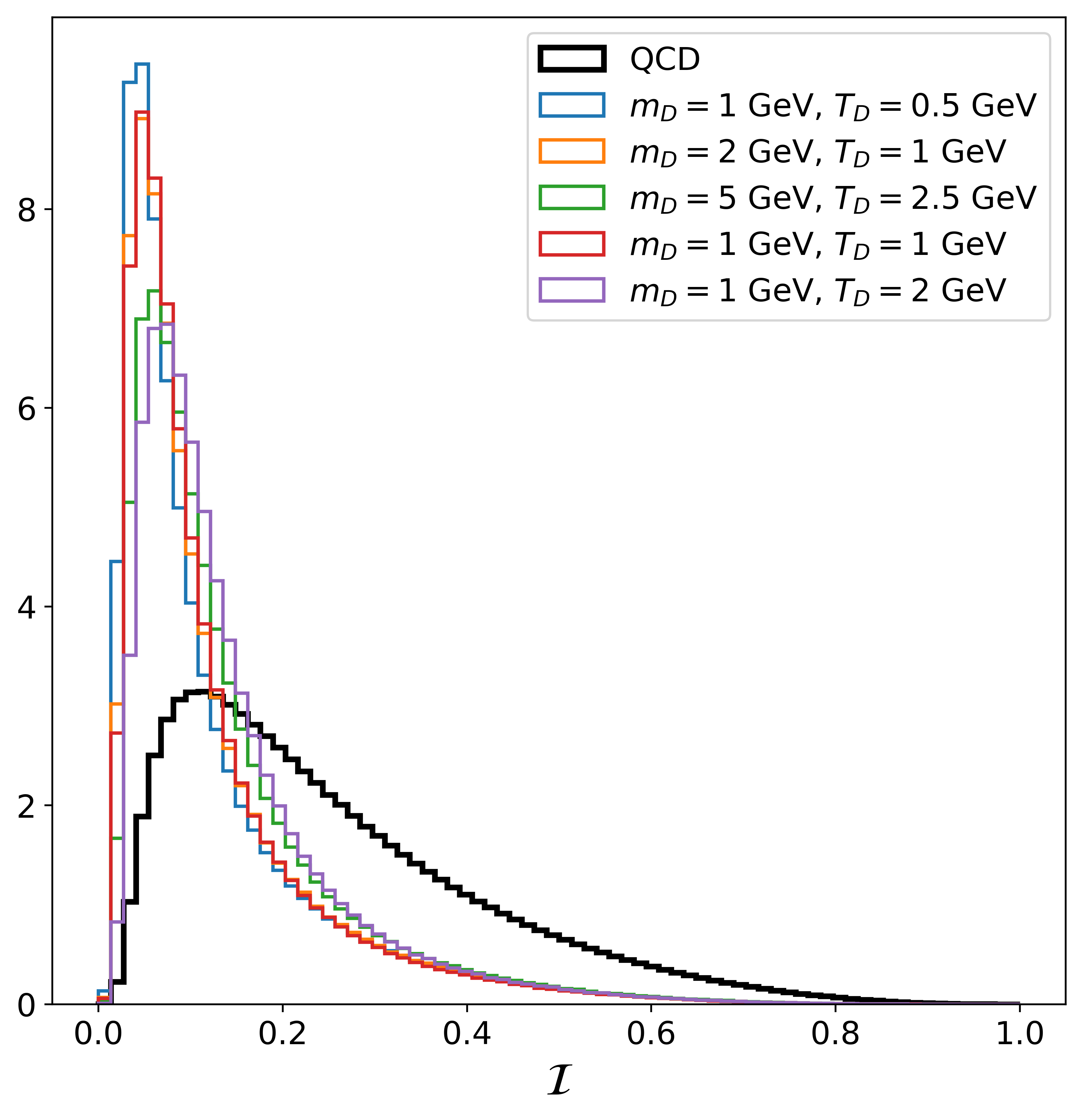}
     \includegraphics[width=0.32\textwidth]{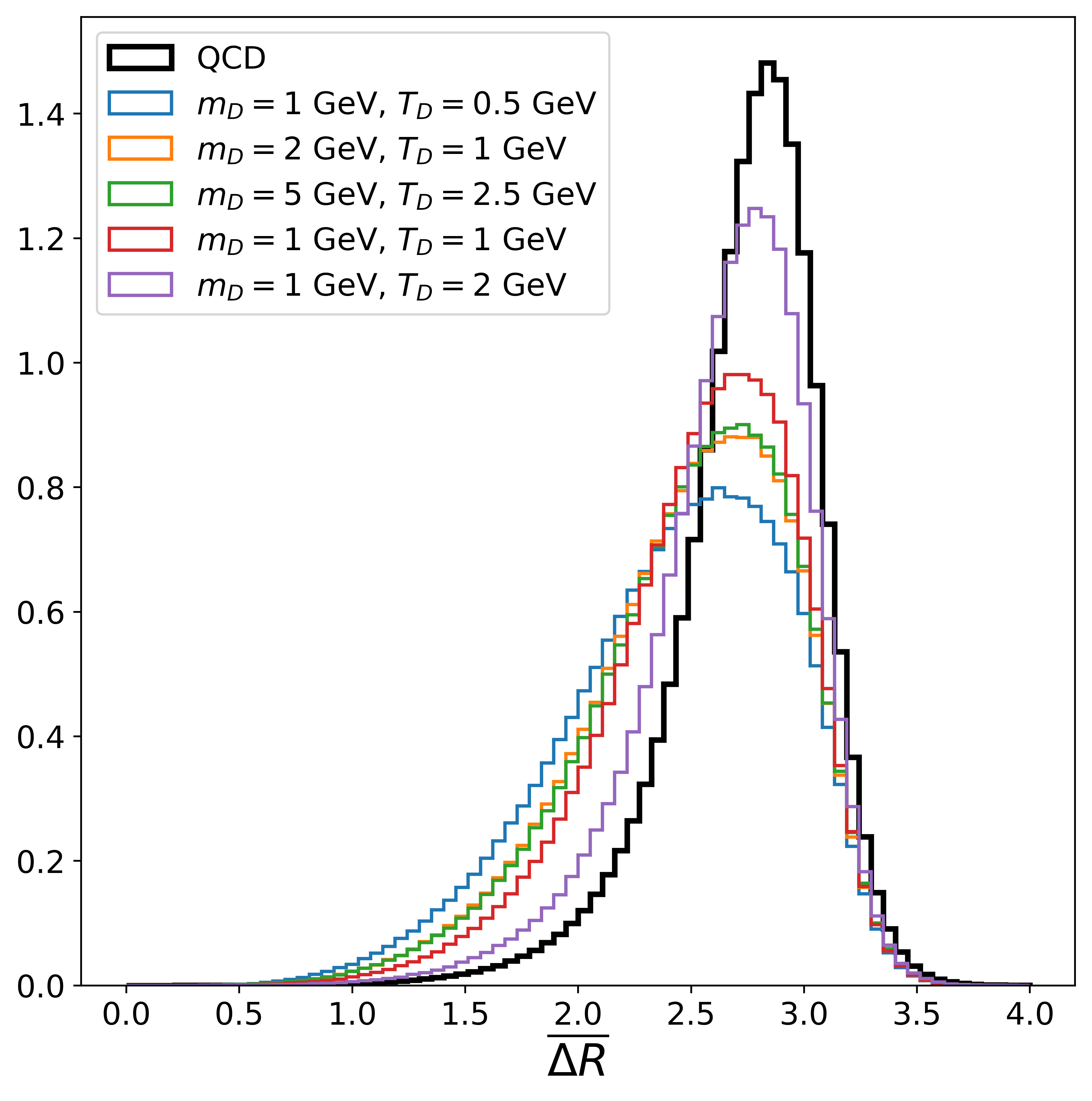}
     \caption{Comparison of QCD and SUEP distributions of selected observables.}
     \label{fig:observables_qcdandsuep}
\end{figure}

Figure \ref{fig:observables_qcdandsuep} shows distributions of $N_\text{charged}$, $\mathcal{I}$, and $\overline{\Delta R}$ for the QCD background and a variety of SUEP benchmark points after trigger selection. The separation between signal and background is clear, with SUEP having higher multiplicity, more isotropic distribution of tracks, and a significantly wider spread of inter-particle distances. 
To understand how these observables change across SUEP parameter space, we show their average values (across the whole sample after trigger selection) as a function of $m_D$ and $T_D$ in Figure \ref{fig:observables_suep}. The pairwise correlations between each of these observables are included in the Appendix in Figure \ref{fig:observables_correlation}, demonstrating that each of these three variables encode distinct information about each event.

In addition to the trigger requirements, we therefore impose the following pre-selection cuts on all events:
\begin{equation}
    \label{e.baseline}
    N_\text{charged} \geq 70 \ \ \ , \ \ \  \mathcal{I} < 0.07 \ \ \ , \ \ \ \overline{\Delta R} < 3 \ .
\end{equation}
This cut targets the most SUEP-like parts of signal parameter space with low dark Hagedorn temperature and/or low dark hadron mass (see \fref{fig:Cartoon}). All but 2.2\% of the post-trigger background is eliminated, while  leaving  $31.8\%$ of signal for $m_D=0.4$~GeV and $T_{D}=0.4$~GeV.
These requirements are less optimal for larger dark hadron masses or temperatures --- for example, the signal efficiency is only $1.1\%$ for $m_{D}=5$~GeV, $T_{D}=20$~GeV.
However, larger temperatures and masses generally lead to higher-energy final states or separable resonances, and are therefore not the focus of our present analysis.

\begin{figure}[t]
     \includegraphics[width=0.32\textwidth]{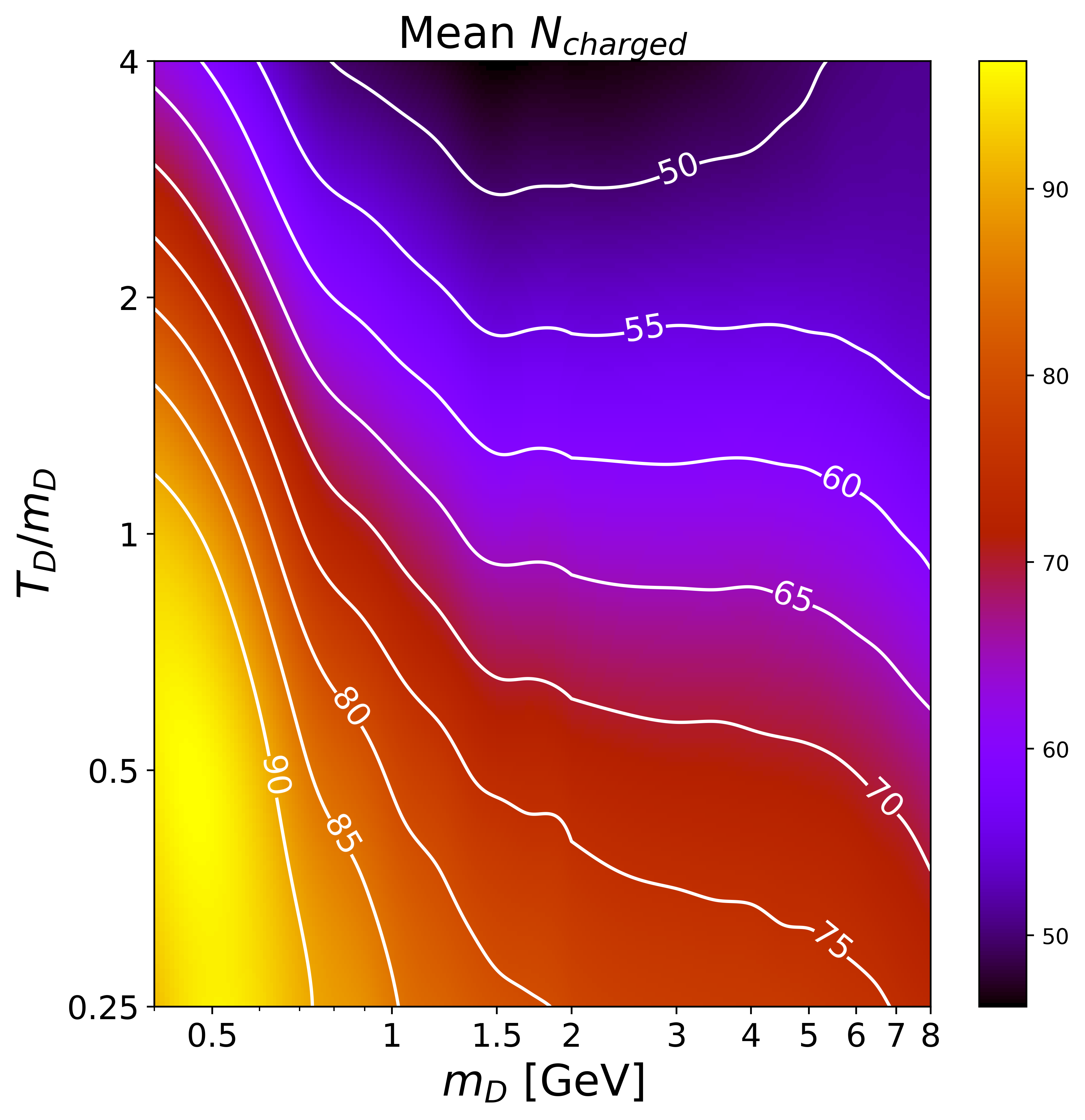}
     \includegraphics[width=0.325\textwidth]{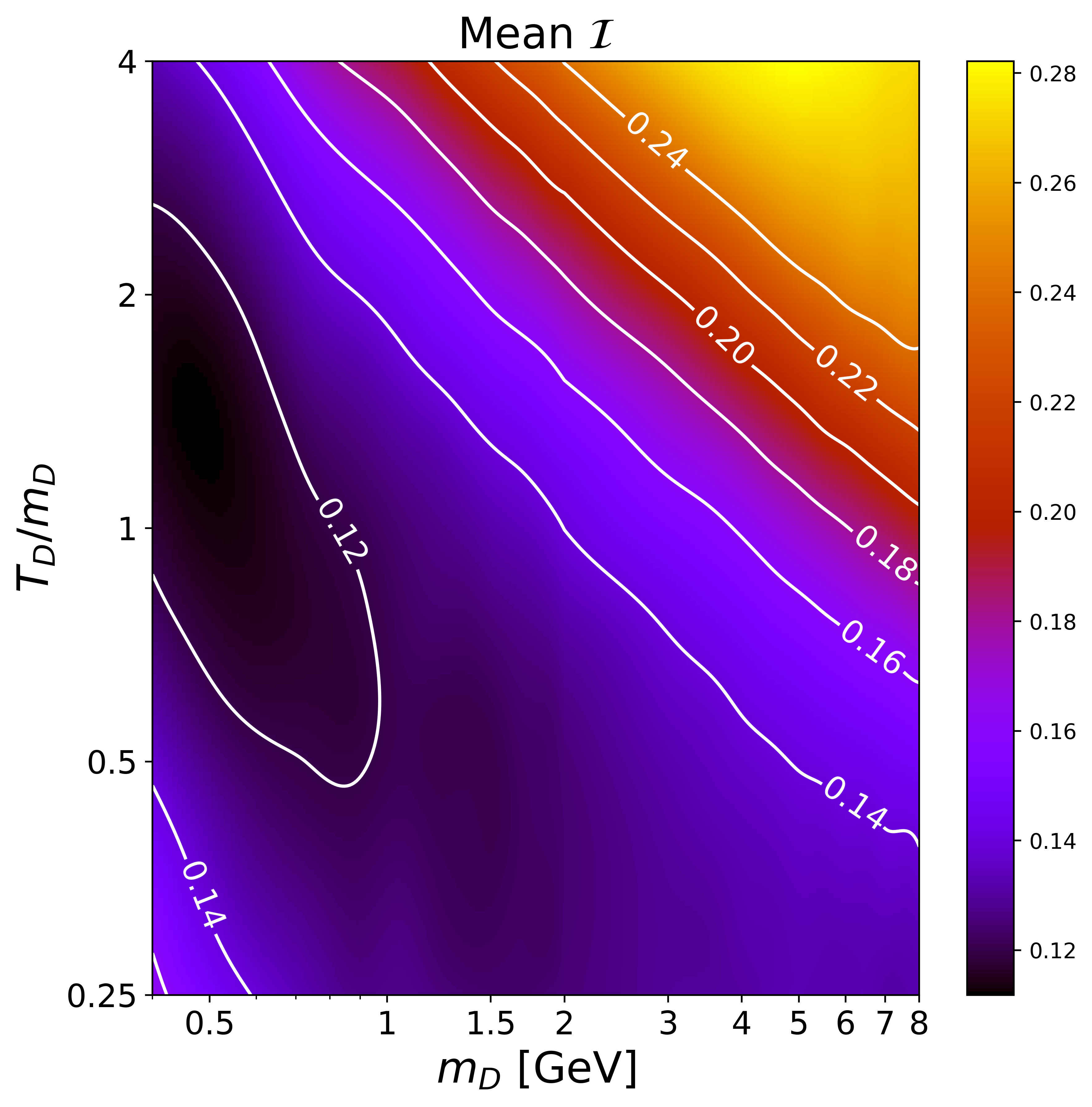}
     \includegraphics[width=0.32\textwidth]{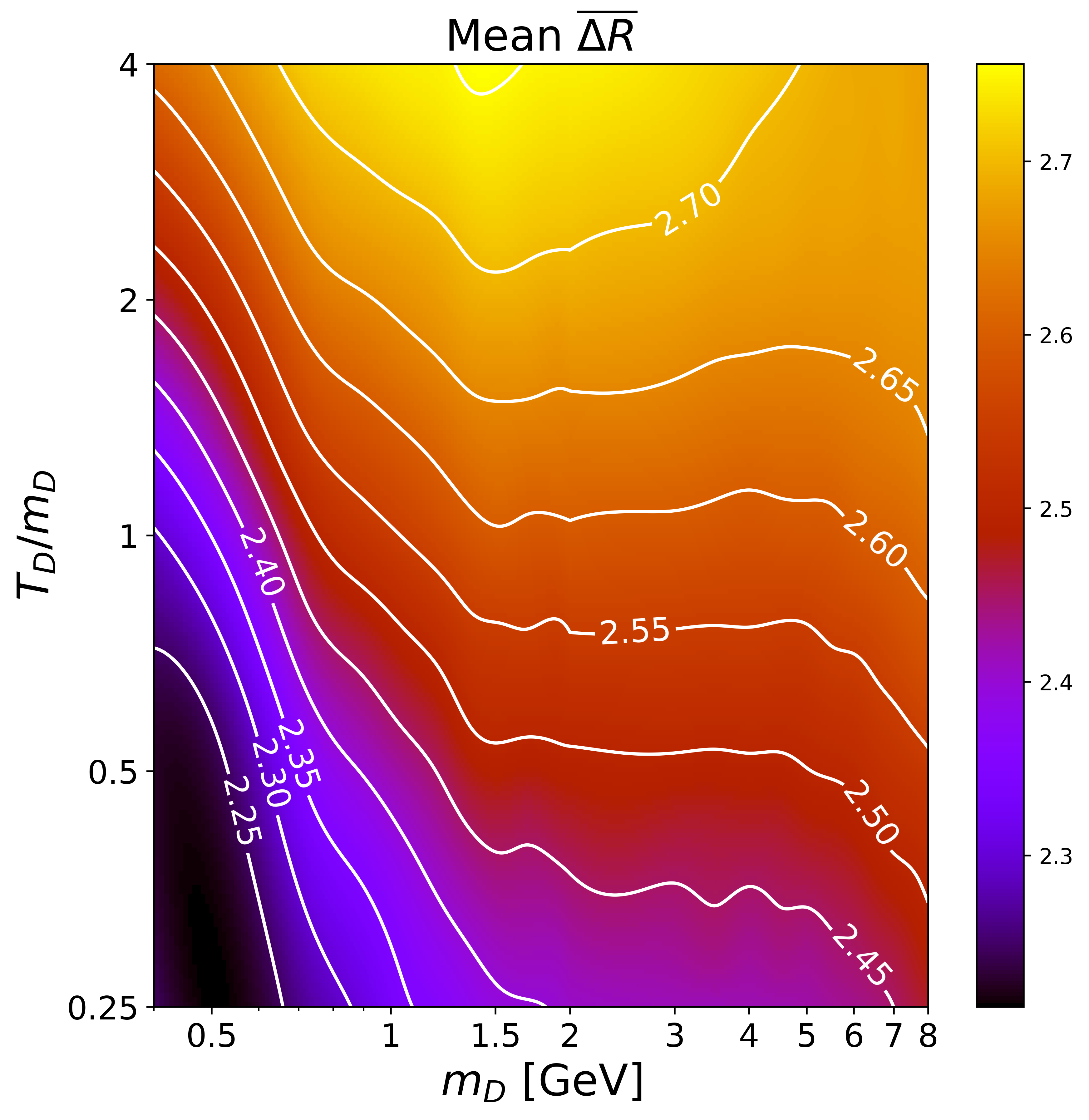}
     \caption{Average values of charged particle multiplicity, event isotropy and mean-interparticle-distance as a function of $m_{D}$ and $T_{D}$ for SUEP. QCD average values: $\langle {N}_\text{charged} \rangle =51$, $\langle{\mathcal{I}} \rangle=0.25$, $\langle \overline{\Delta R} \rangle=2.7$.
     }
     \label{fig:observables_suep}
\end{figure}
\bigskip

We now consider three options for SUEP searches at the HL-LHC, using events which pass the baseline pre-selection as a starting point:
\begin{enumerate}
    \item The simplest strategy is a cut-and-count analysis using high-level observables. It will serve as a baseline for more sophisticated machine-learning techniques, and can be implemented very easily and effectively with a stricter cut on $\overline{\Delta R}$ compared to \eref{e.baseline}. Varying the cut threshold yields a significance improvement curve (SIC) of the signal efficiency vs the background efficiency for each point in signal parameter space, which we will then be able to compare to results using machine learning methods. As we will see, this already yields very promising SUEP sensitivities.
    
    \item A supervised ML-classifier requires detailed knowledge of the signal, since it is trained on signal and background. This makes supervised techniques unlikely to be a realistic analysis candidate for broad SUEP searches. However, we perform a simple supervised study in \sref{sec:supervised} to demonstrate the best-case scenario for SUEP sensitivity if the signal was very well understood.
    
    \item In \sref{sec:unsupervised}, we use an unsupervised autoencoder trained only on the background as an anomaly detector to improve on the sensitivity of the cut-and-count analysis. This is likely to be a realistic analysis candidate since it can be performed using data-driven background estimation techniques without precise knowledge of the signal.
\end{enumerate}

\subsection{Supervised ML-Classifier}\label{sec:supervised}

While a supervised classifier is explicitly dependent on the characteristics of the signal (and background) simulation used in its training, which we cannot trust in detail, it can still provide a useful comparison to an unsupervised network, and give an indication of how sensitive a search based on similar methods could be with improved modelling of the SUEP signal. An additional limitation is that even with reliable signal simulation, the parameters of the real SUEP are unknown, and a supervised network trained to recognize SUEP with one set of parameters may fail if the true parameters change. As we will see, this is indeed the case. 

The supervised network architecture we choose is a dynamical graph convolutional neural network~\cite{wang2019dynamic,Bernreuther:2021gds}. We implement the network in PyTorch. 
The input feature representation for both the supervised and unsupervised networks is the interparticle distance matrix, with the redundant left-lower half set to zero and track $p_T$ information added to the diagonal:
\begin{equation}
    \label{e.inputfeatures}
    \Delta \tilde R_{ij} \equiv \left\{
    \begin{array}{lll}
    \Delta R_{ij} & & i > j\\
    p_{T,i}/\mathrm{GeV} \ \ \   & \mathrm{for} \ \ \  & i = j\\
    0 &  & i < j
    \end{array} \right.
\end{equation}
All events are required to have at least $N_\text{charged} \geq 70$ tracks, and all events are truncated to keep only the $70$ highest-$p_{T}$ tracks to ensure each event has the same dimensionality in the analysis.

The input matrix $\Delta R_{ij}$ is used to generate graph edges between each particle (node) and its $k=7$ nearest neighbours in $\Delta R$ space. The node features for each particle are the 70-dimensional vector of $\Delta R$ distances to all other particles in the event. The graph network has two EdgeConv blocks, each comprising a three-layer perceptron with leaky ReLU activation~\cite{Qu:2019gqs}. The EdgeConv operation updates the node features $x_{i}$ of each particle as 
\begin{equation}
    \vec{x}_{i}^{'} = \frac{1}{k}\sum_{j=1}^{k}\vec{h}_{\theta}(\vec{x}_{i},\vec{x}_{i}-\vec{x}_{j}) \; ,
\end{equation}
where $k$ is the number of neighbours assigned to each node, and $h_{\theta}$ is a non-linear function of learned parameters $\theta$, implemented as a three-layer perceptron. 
The graph edges are re-computed between the first and second blocks using the Euclidean distance between the feature vectors of each node to determine its $14$ nearest neighbours. The first block has feature dimension 64, while the second block has feature dimension 32. Batch normalization follows each layer. The output of the graph layers is averaged, then passed through two fully connected layers, first expanding to dimension 128, then down to output dimension 2. The loss function is the cross-entropy loss between the output and the true class label of each event. 
\begin{equation}
    \mathcal{L}(x,\text{class}) = -\log \frac{x[\text{class}]}{\sum_{j}x[j]}
\end{equation}

To test how the performance of the model depends on the choice of training parameters, we train twelve neural networks on twelve different choices of $(m_D, T_D/m_D)$, with $m_{D} = 0.5, 1, 2 \gev$ and $T_{D}/m_D = 0.5, 1, 2, 4$, and evaluate their effectiveness over the whole signal parameter space.
We also evaluate the efficacy of a `cocktail approach'~\cite{Aguilar-Saavedra:2017rzt,Knapp:2020dde,Bernreuther:2021gds,Baldi:2016fzo} by training on a mixed sample including signal events from each of the twelve parameter choices. A conditional training~\cite{Louppe:2016ylz,CMS:2019dqq} on the signal parameters would be a similar approach.
Each network is trained for 10 epochs with a decaying learning rate. Longer training periods were tested and found to be unnecessary. Loss values for each test sample event are obtained for the model realizations of the last 5 training epochs before being averaged.

\subsection{Unsupervised Autoencoder}\label{sec:unsupervised}

Supervised ML-classification is an extremely powerful analysis tool, but it only works for signals with well-defined and universal features and  corresponding precision simulations. At the LHC, this is not always the case, and SUEP with its toy shower is a perfect example for a more broadly defined signal. Here we prefer not to train a classifier on signal simulations. Instead, we employ anomaly detection methods, where we train a network only on the well-understood background dataset, so that it can flag events that are anomalous in comparison. We use an autoencoder~\cite{Rumelhart1986,Heimel:2018mkt,Farina:2018fyg} and train it on data without class labels, with a loss function that incentivizes its output to be as close as possible to the input. The intermediate network layers have a restricted number of nodes compared to the dimension of the input and output, forcing the network to compress the information in the input, and then decompress it to recover the output. The principle of the autoencoder's use as an anomaly detector is that it should fail to accurately reconstruct events that are anomalous compared to the dataset it was trained on. A high reconstruction loss flags an event as being potentially anomalous, or in collider physics parlance, a signal (SUEP) candidate event.

The autoencoder uses the same modified $\Delta \tilde{R}_{ij}$ matrix input as used by the supervised network, see \eref{e.inputfeatures}. Other representations were tested, including the matrices of both $k_{T}$ and anti-$k_{T}$ distances $d_{ij} = \min(k_{ti}^{\pm 2},k_{tj}^{\pm 2})(\Delta \tilde{R}_{ij})^{2}/R$ between particles~\cite{Cacciari:2008gp}, the high-level observables used in the pre-selection cuts, as well as the raw $p_{T},\phi,\eta$ values for each particle; all yielded results that were much less useful than the analysis we present here. This emphasizes the importance of choosing the correct input representation over sophisticated network architecture for SUEP searches. 

The matrix  $\Delta \tilde{R}_{ij}$ is flattened into a vector of length $N_\text{charged}^{2}$ and fed into the autoencoder. The neural network comprises five fully connected layers, with the number of nodes decreasing to the bottleneck size in the third layer, then increasing back to $N_\text{charged}^{2}=4900$. 
An alternative 3-layer network performs only slightly worse. For the bottleneck size we find  that larger bottlenecks consistently lead to better performance than smaller ones, so we use $N_\text{bottleneck}=1000$. Each layer of the network has a leaky ReLU activation with slope -0.2 for negative values of x, except the final layer which has a ReLU activation. 
The loss function of the network measures the difference between the network's input and output as 
\begin{equation}
\mathcal{L}(y^{\text{in}},y^{\text{out}}) = \frac{1}{N_\text{charged}}\sum_{i}\frac{|y^{\text{out}}_{i}-\sigma(y^{\text{in}}_{i})|^{m}}{|\sigma(y^{\text{in}}_{i})|^{n}}
\end{equation} 
where $\sigma(x)=1/(1+e^{-x})$. Among different values of $m$ and $n$, starting with the usual mean squared error $m=2,n=0$, the best-performing is $m=3,n=0$. The sigmoid normalization of the input is essential to the network's success. Without it, the autoencoder encodes SUEP events with slightly \emph{lower} loss than the QCD background on which it was trained, and completely fails to identify anomalous events. We hypothesize that this is because, unlike many experimental signatures, SUEP is less complex than its QCD background, and it has smaller values of $\Delta R$ than QCD. This complexity bias has been noted before~\cite{Heimel:2018mkt,Dillon:2021nxw,Finke:2021sdf}. The sigmoid function reduces sensitivity to large values of $\Delta R$ and $p_{T}$ by mapping them to values very close to 1, while remaining approximately linear for small input values, but offset to a minimum value of 0.5. These effects make it easier to accurately reconstruct QCD events while enhancing the network's sensitivity to deviations from the input on the SUEP events with characteristically smaller absolute values of the input features. 

Out of $8.8\times 10^{5}$ background Monte Carlo events that pass the pre-selection cuts, $2.4\times 10^{5}$ are used for training, $5\times 10^{4}$ for validation when tuning network hyperparameters, and $5.9\times 10^{5}$ for testing. The number of signal events that pass the cuts varies with $m_{D}$ and $T_{D}$, but generally a few $\times 10^{4}$ events remain at each parameter point to be used for testing. SUEP events with $m_{D}=1$ GeV, $T_{D}=0.5$ GeV are used for validation purposes.  

The network is trained for 15 epochs with a decaying learning rate. Longer training periods were tested and found to be unnecessary. Loss values for each test sample event are obtained for the model realizations of the last 5 training epochs before being averaged.

Other architectures were investigated, including variational autoencoders~\cite{kingma2014autoencoding} and a graph convolutional autoencoder utilizing the same EdgeConv operations as the supervised network, which itself led to the use of the $\Delta \tilde R_{ij}$ event representation. Interestingly, the simpler, fully connected architecture consistently delivered much better background rejection than any of the more sophisticated graph networks in the unsupervised approach.

\subsection{Data-driven Background Estimation}
\label{ss.datadriven}

Following the logic of this section further, we briefly describe how a data-driven background sample could be derived for use in a realistic experimental analysis based on our study.
The total background cross-section can be measured directly (and compared against simulation), since the leptons+QCD signal region is completely background dominated for SUEP production in exotic Higgs decays. The background efficiency of the classifier, whether it is based on cuts, an unsupervised neural network, or a supervised one, can be estimated by evaluating it on a control region.

A variety of control regions are possible, but perhaps the most promising is defined by replacing the lepton criterion by a mono-photon criterion. 
Such a sample should be free of signal contamination, and its hadronic content is extremely similar to the hadronic background to our SUEP search, since it is unconstrained in its detailed production channel except that it recoils off a hard electroweak boson. 
The QCD distributions in this mono-photon-plus-jets sample should  therefore closely match those in the $Z$+jets signal region, especially if the control region sample is reweighted to match the shape of the photon $p_T$ spectrum to the dilepton $p_T$ spectrum in the $Z$+jets signal sample. 
A variant of this approach can likely be adapted to estimate the background in the $W$+jets channel as well, by using simulation to compute a transfer function from  $p_{T,W}$ to $p_{T,\ell}$ and applying it to $p_{T,\gamma}$ in the control region before applying the $p_{T,\ell}$ reweighing. 

Based on the assumption that such a strategy can be implemented, we therefore will use benchmark estimates of $1\%$ and $10\%$ for the systematic background uncertainty in our analysis to estimate the final physics reach.

%%%%%%%%%%%%%%%%%%%%%%%%%%%%%%
%%%%%%%%%%%%%%%%%%%%%%%%%%%%%%
%%%%%%%%%%%%%%%%%%%%%%%%%%%%%%
%%%%%%%%%%%%%%%%%%%%%%%%%%%%%%
%%%%%%%%%%%%%%%%%%%%%%%%%%%%%%

\section{Results}
\label{sec:results}

\begin{figure}
    \centering
    \begin{tabular}{cc}
    
    \includegraphics[width=0.45\textwidth]{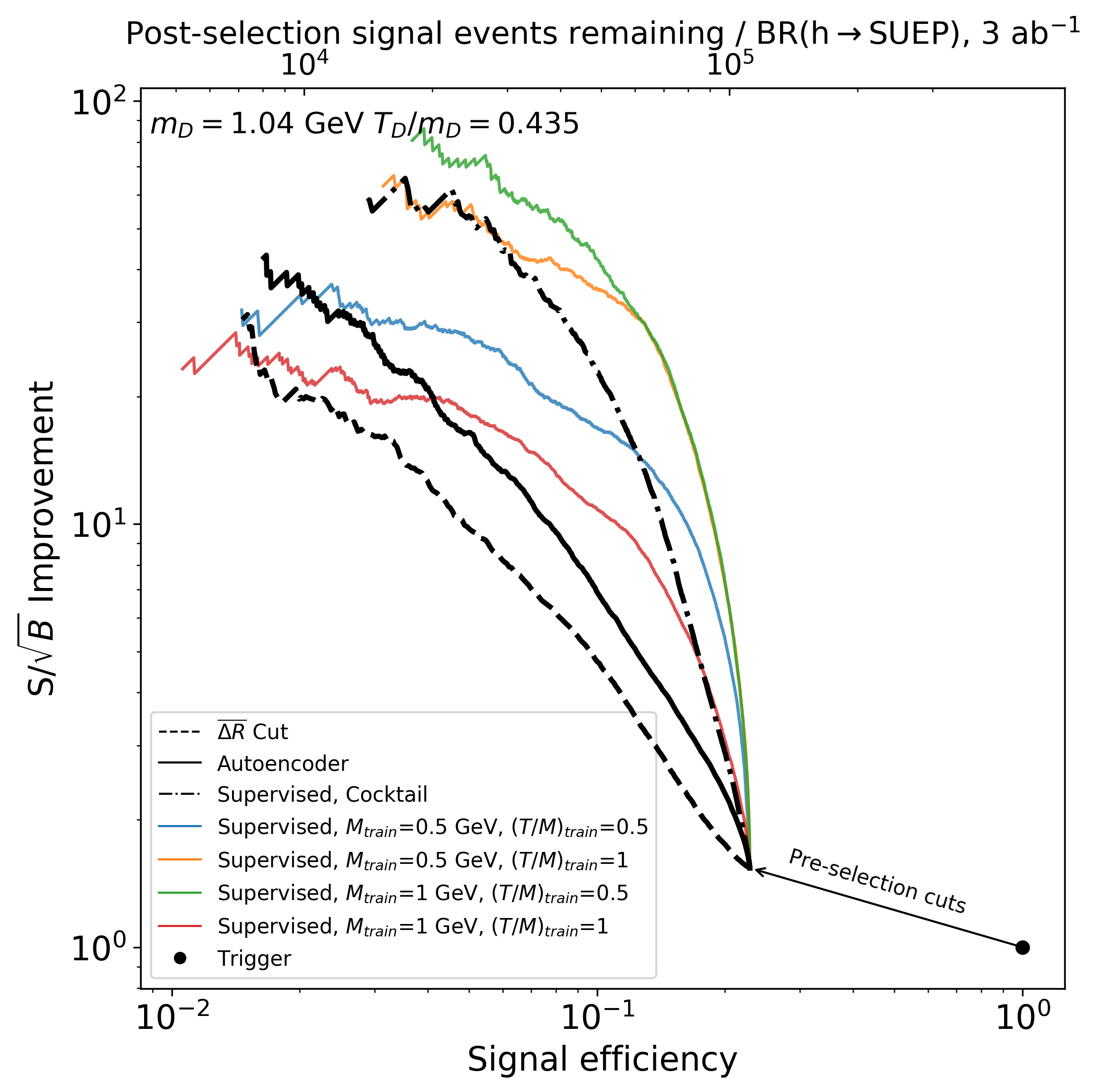}
        
    &
    \includegraphics[width=0.45\textwidth]{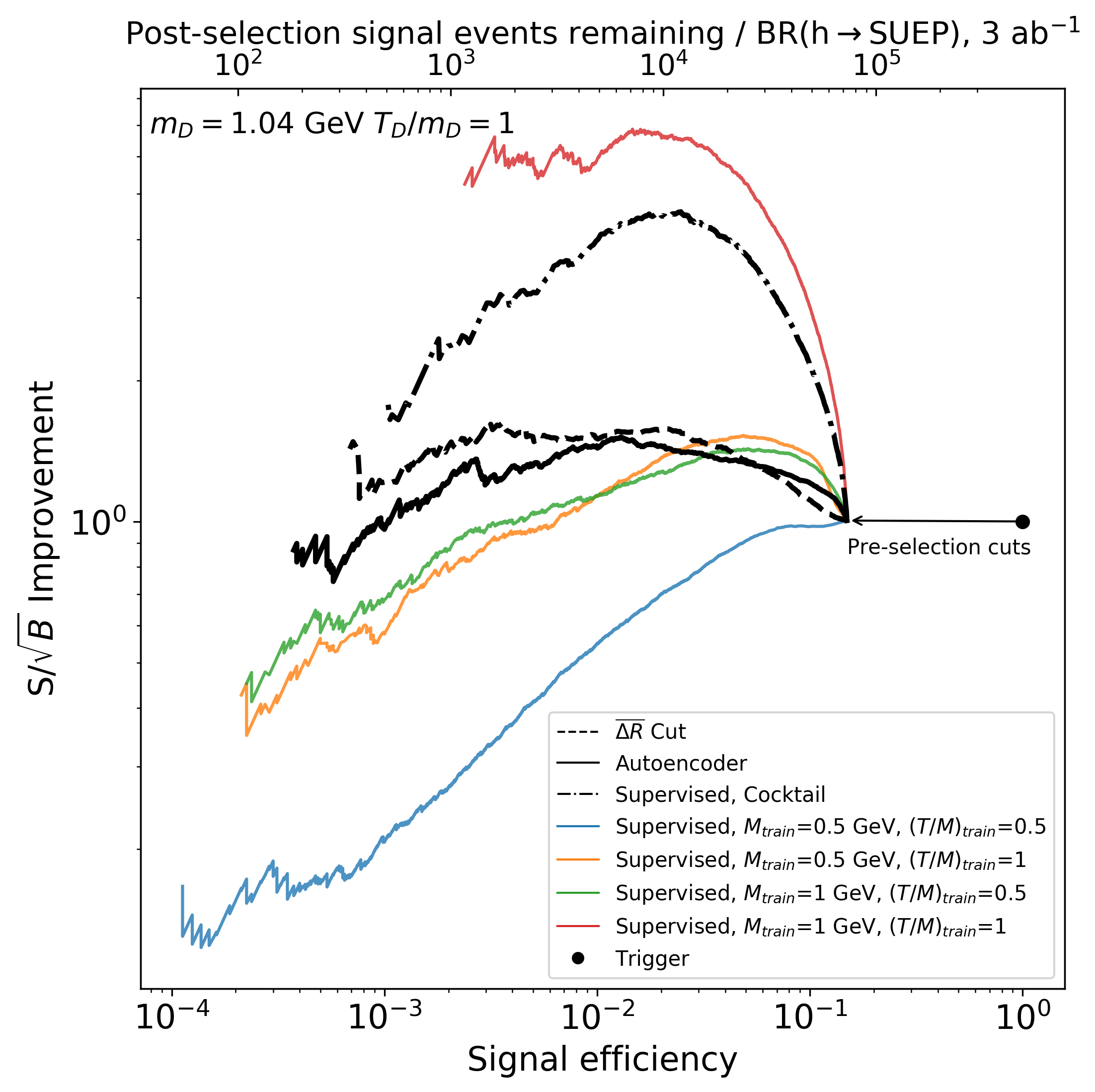}
           
    \\
    
    \includegraphics[width=0.45\textwidth]{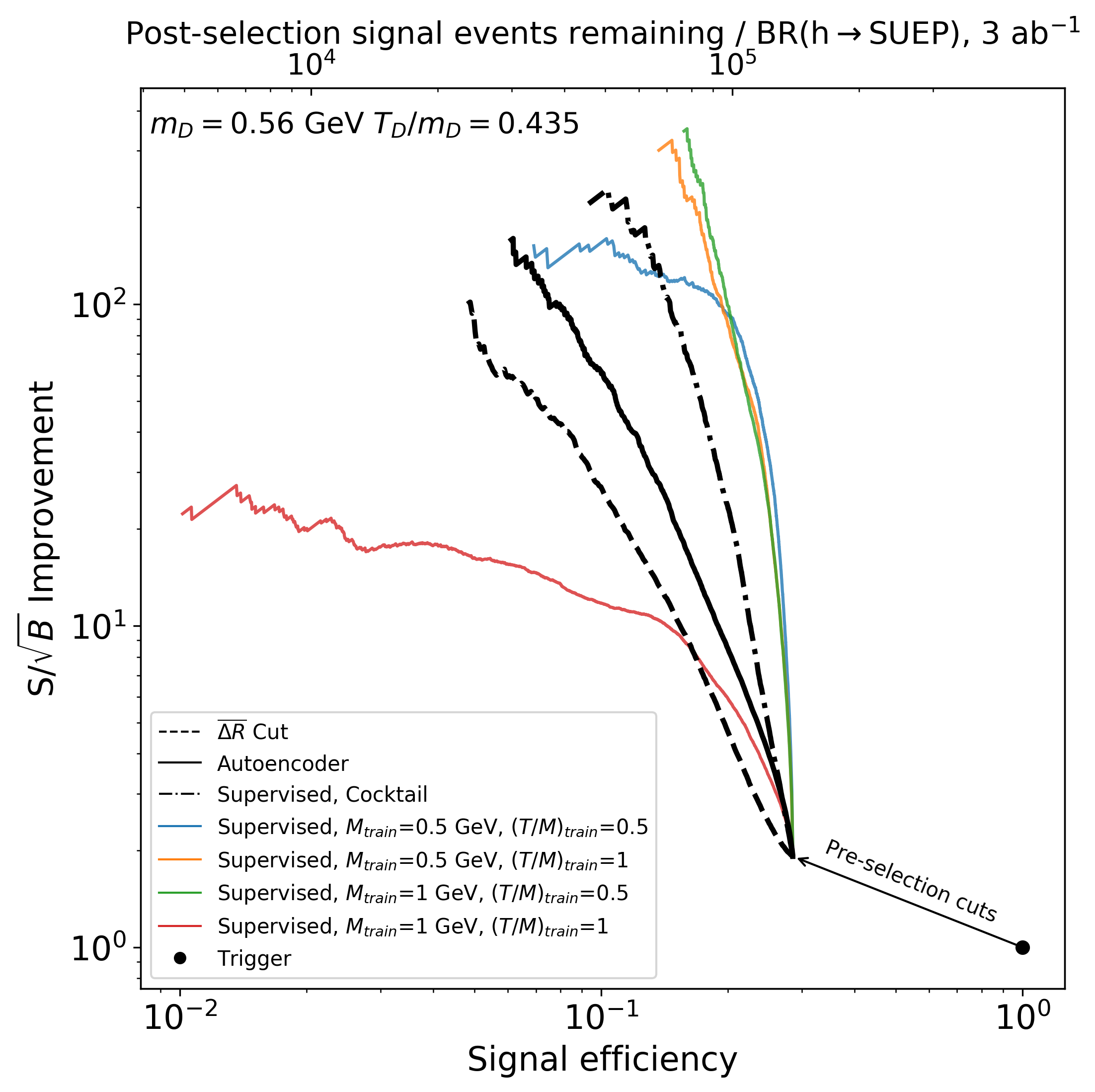}
    
   &
    
    \includegraphics[width=0.45\textwidth]{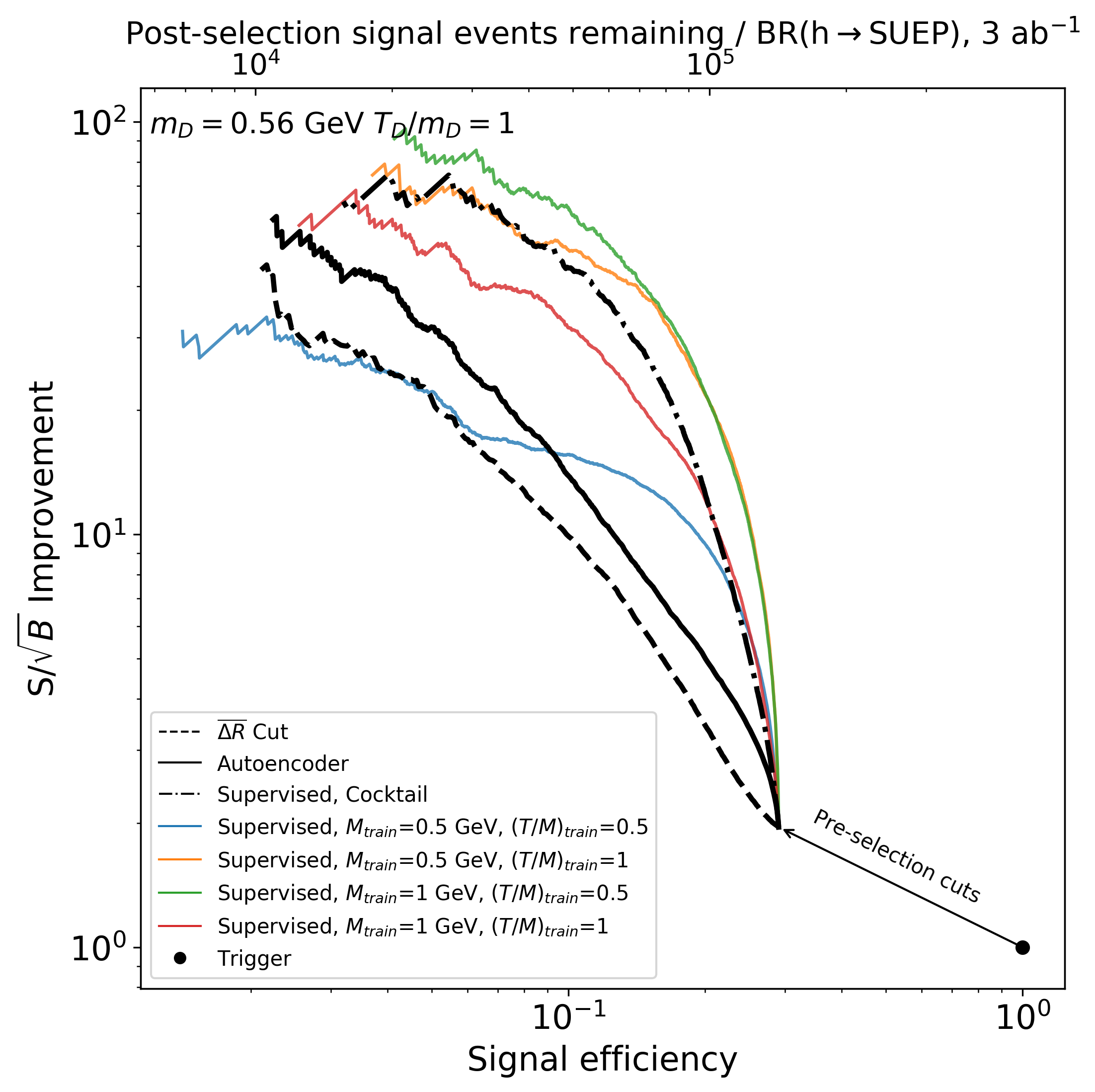}
    
    \end{tabular}
    \caption{Some examples of Significance Improvement Curves (SIC), relative to trigger selection, of the autoencoder (black solid), cut on $\overline{\Delta R}$ (black dashed), and supervised graph networks (black dash-dotted and colored) for SUEP with test parameters of $(m_{D}, T_D/m_D) =$
    (1.04, 0.435), 
    (1.04, 1), 
    (0.56, 0.435), 
    (0.56, 1).
    For the autoencoder, the peak sensitivity improvement we are able to probe reliably with the statistics of our Monte Carlo samples (see text) corresponds to signal and background efficiencies of 
    $(1.6 \times 10^{-2}, 1\times10^{-7})$,
    $(3.7\times10^{-4}, 1.5\times10^{-7})$,
    $(6.1\times 10^{-2},1\times10^{-7})$ 
    and 
    $(2.2 \times 10^{-2}, 1\times10^{-7})$ relative to trigger selection, respectively.
    }
    \label{fig:ROC_comparison}
\end{figure}

\begin{figure}
    \centering
    \begin{tabular}{cc}
    
    \includegraphics[width=0.45\textwidth]{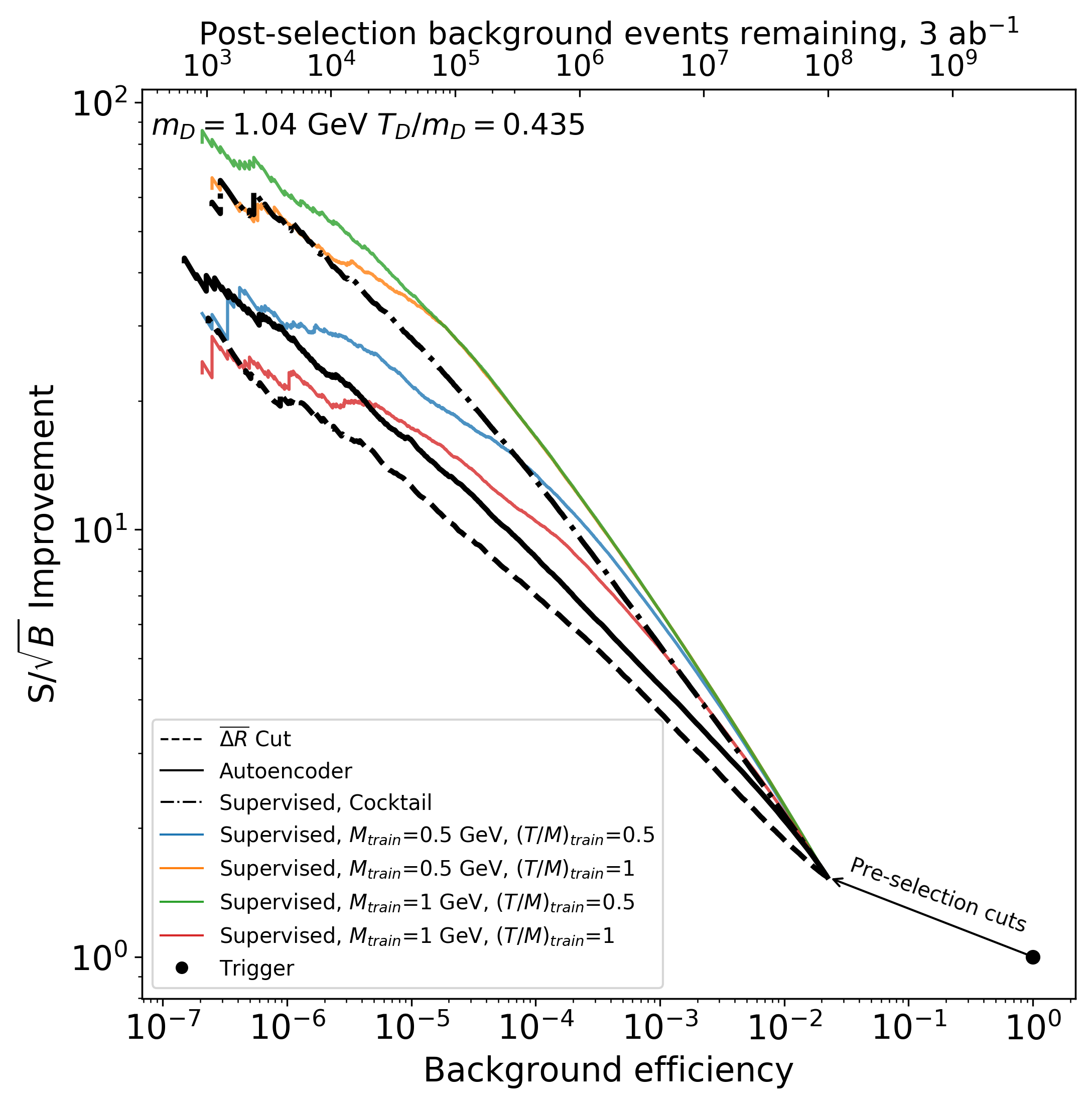}
        
    &
    \includegraphics[width=0.45\textwidth]{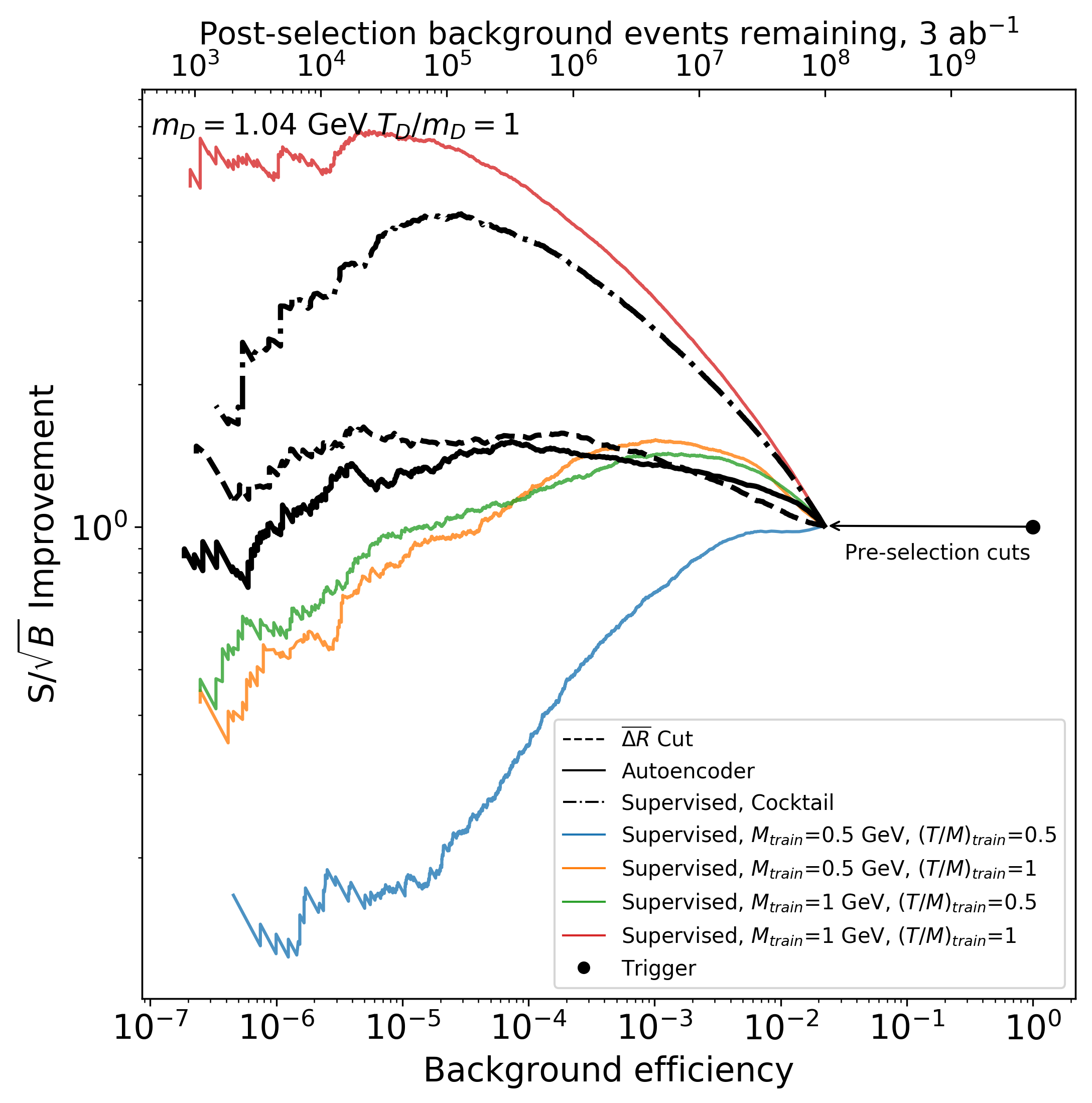}
           
    \\
    
    \includegraphics[width=0.45\textwidth]{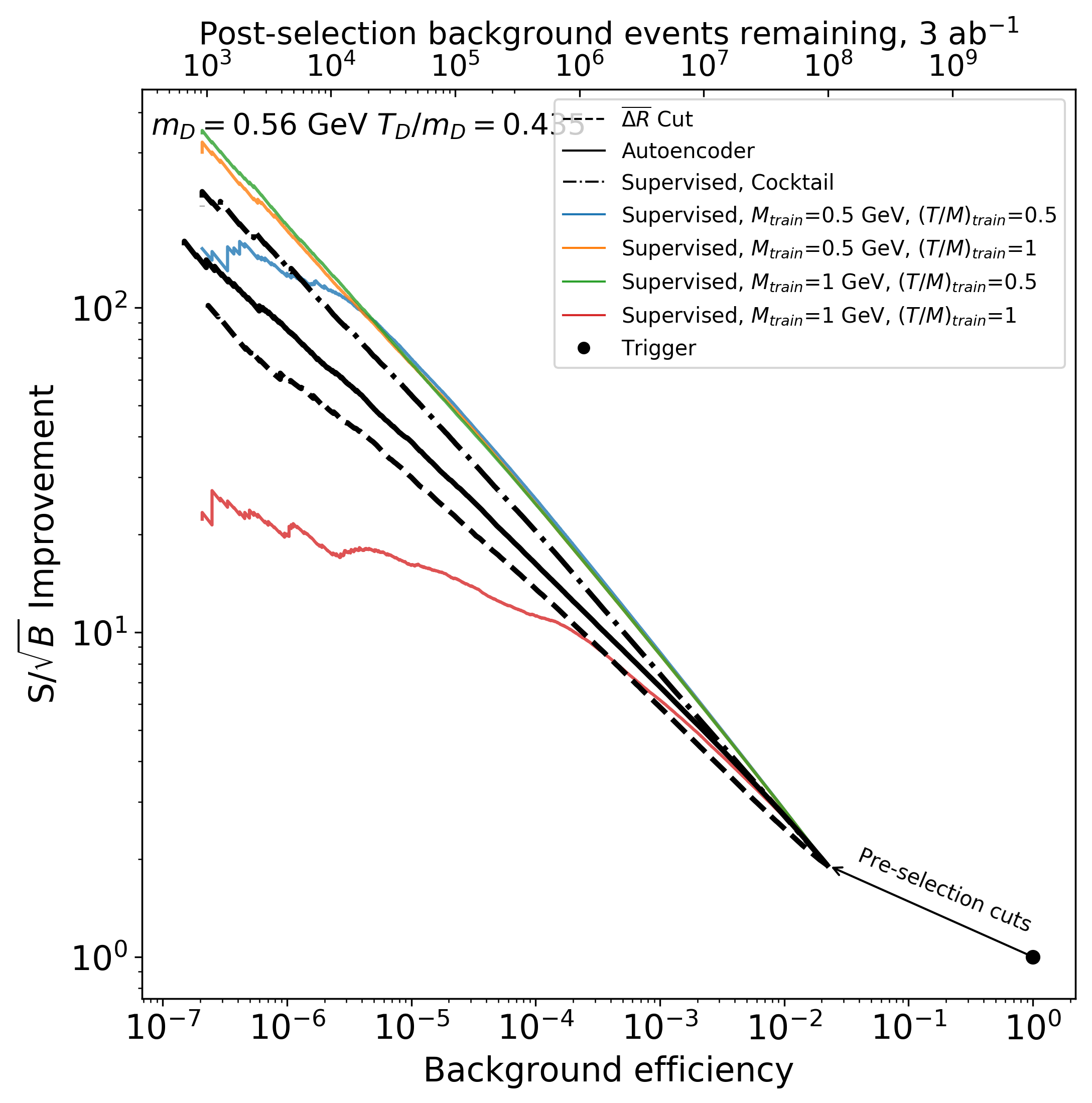}
    
   &
    
    \includegraphics[width=0.45\textwidth]{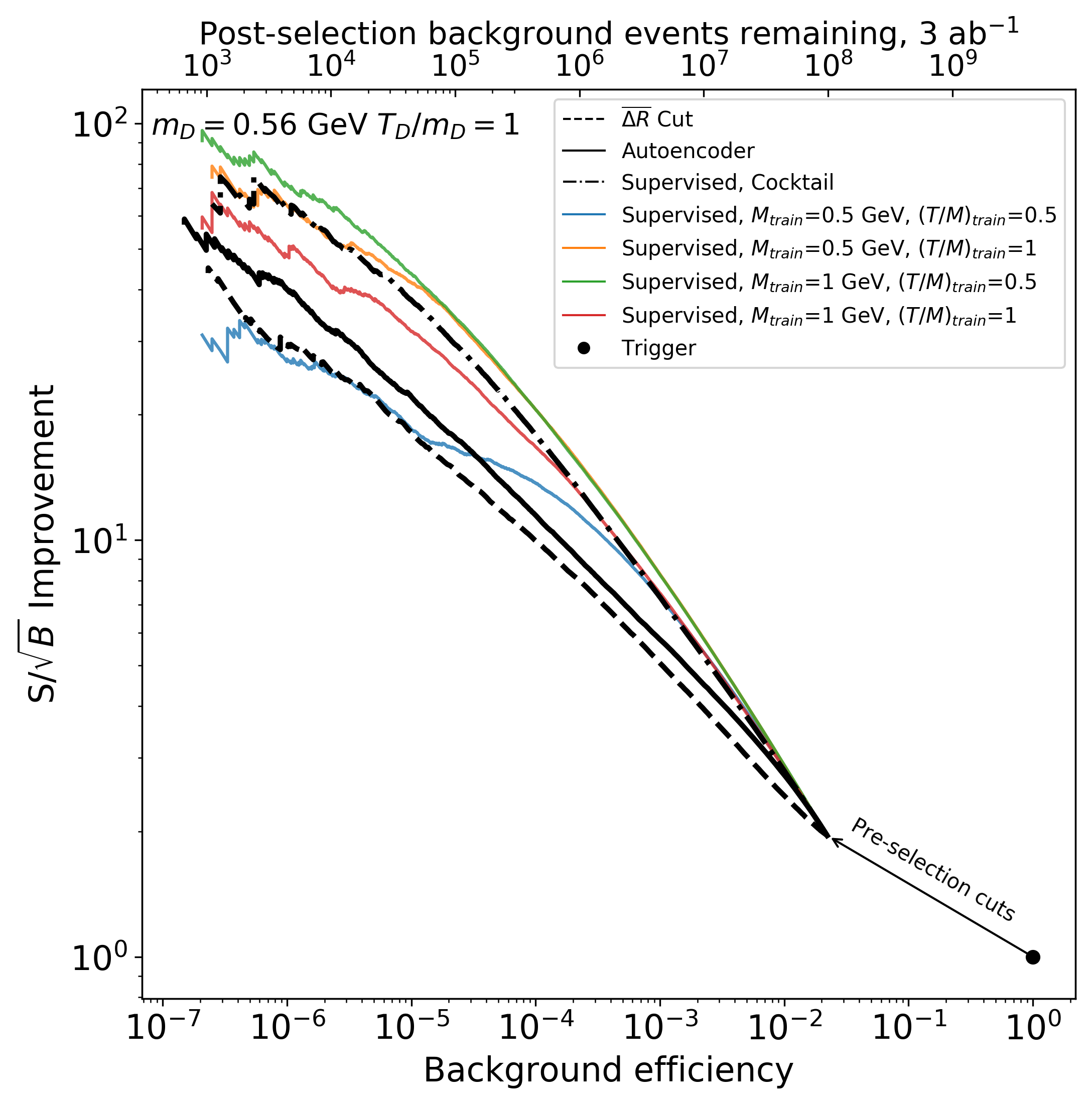}
    
    \end{tabular}
    \caption{Same SIC curves as \fref{fig:ROC_comparison_alt}, but with background efficiency on the horizontal axis. This demonstrates that with the full HL-LHC dataset, significantly harder cuts could increase sensitivity beyond the limitations of what we can demonstrate with our simulated background dataset. 
    }
    \label{fig:ROC_comparison_alt}
\end{figure}

Figure~\ref{fig:ROC_comparison} shows significance improvement curves
for the cut on $\overline{\Delta R}$, the autoencoder, and a selection of supervised networks trained on signal samples with a variety of different dark hadron masses and temperatures $m_\text{train}$ and $T_\text{train}$. The horizontal axis shows either signal efficiency (after triggering) or number of remaining signal events at the HL-LHC  as the threshold is varied. The vertical axis shows signal significance improvement relative to the trigger sample. 
From these specific examples, a few general features are apparent:
\begin{itemize}
    \item The autoencoder consistently outperforms the simple $\overline{\Delta R}$ cut significantly;
    \item The supervised networks outperform the autoencoder for signal parameters close to their training values, but can perform much worse for different parameters;
    \item Our analysis is not optimized for larger dark hadron masses and temperatures;
    \item For lower dark hadron masses or temperatures, both the cut-and-count and autoencoder analysis strategies are very powerful, yielding orders of magnitude improvement in signal significance compared to the  baseline preselection cuts of \eref{e.baseline}.
\end{itemize}
In fact, the statistics of our QCD background sample is insufficient to capture the true power of our analysis techniques. We would expect the significance improvement to increase as the signal efficiency decreases, but only until the SI curve turns around when the cut becomes too harsh. 
This is clearly the case in the top-right panel of \fref{fig:ROC_comparison}, for signal parameters to which our analysis is not optimized. However, for the three other examples, we never reach this turn-over point before running out of simulated background events. It is not even clear if this turn-over is reached with the full statistics of the HL-LHC (100 times greater than our simulated background sample).

To understand how much harder we might be able to cut on the background, \fref{fig:ROC_comparison_alt} shows the same SIC curves but with background efficiency on the vertical axis. 
Naively extrapolating, we can anticipate that a realistic autoencoder search with a fully data-driven background sample at the HL-LHC might be able to reach the very-low background regime while retaining enough signal to probe $\mathrm{Br}(h \to \ \mathrm{SUEP})$ as much as an order of magnitude smaller than the sensitivity estimates we can rigorously derive here.

As a result, the reach projections we present in this paper will be very conservative. Furthermore, the limited statistics of our background sample means that reach projections will be very similar for the three analysis methods despite their obvious differences in performance in Figs.~\ref{fig:ROC_comparison} and~\ref{fig:ROC_comparison_alt}. 
It is therefore important to additionally evaluate our classifiers using a somewhat orthogonal metric.\bigskip

\begin{figure}
    \centering
    \includegraphics[width=0.9\textwidth]{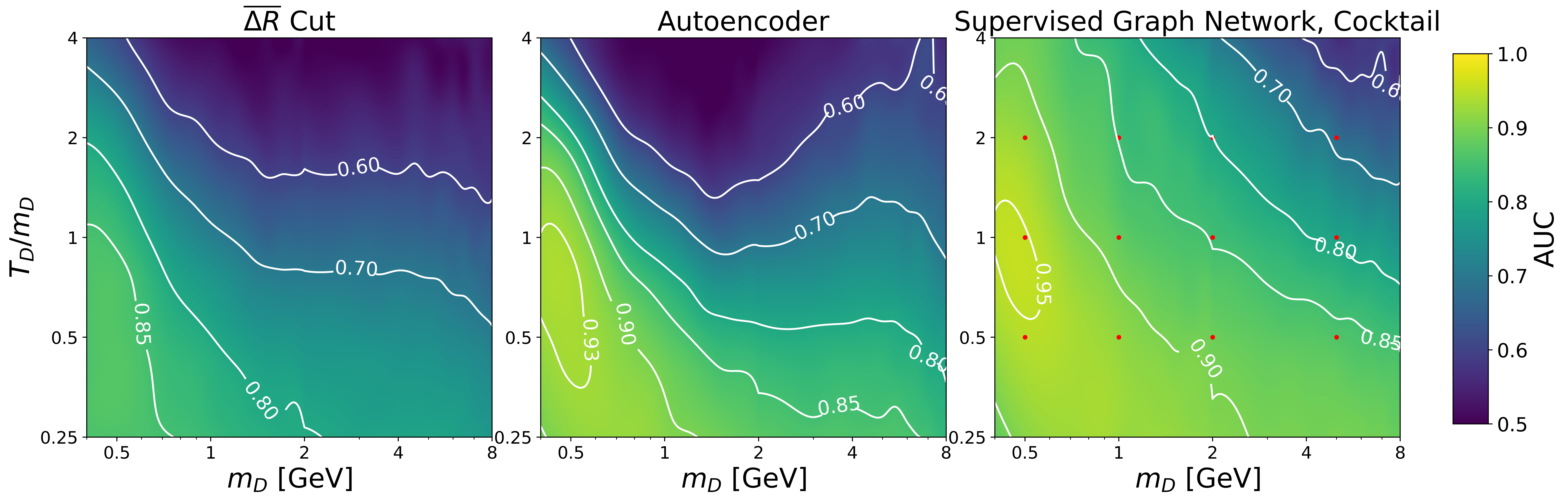}
    \\
    (a)\phantom{blablablablablablabla}(b)\phantom{blablablablablablabla}(c)\phantom{bla}
    \caption{
    AUC for (a) cut on $\overline{\Delta R}$, (b) fully connected autoencoder, (c) supervised graph network trained using cocktail of signal parameter choices (training parameters indicated with red dots). These plots illustrate the significant performance improvements of the autoencoder relative to the simple $\overline{\Delta R}$ cut, and of the supervised cocktail approach relative to the autoencoder.
    }
    \label{fig:auc}
\end{figure}

The area under the curve (AUC) of the signal efficiency as a function of background rejection can be computed 
as a function of the number of events kept after the baseline preselection cuts, see \eref{e.baseline}. This metric is standard to measure the performance classifiers independently of the choice of threshold. \fref{fig:auc} shows the AUC achieved by the cut-based classifier, the fully connected autoencoder, and the supervised graph network trained on a cocktail of signal events, where for the latter the signal parameters are indicated with red dots in the $(m_D, T_D/m_D)$-plane). Unsurprisingly, the highest AUC values are achieved at low $m_{D}$ or $T_{D}/m_D$, with the supervised graph network modestly outperforming the autoencoder, which outperforms the simple cut, across the SUEP parameter space. The performance of supervised networks trained on single signal parameter points are presented in the Appendix.

\begin{figure}
    \centering
    \includegraphics[width=0.9\textwidth]{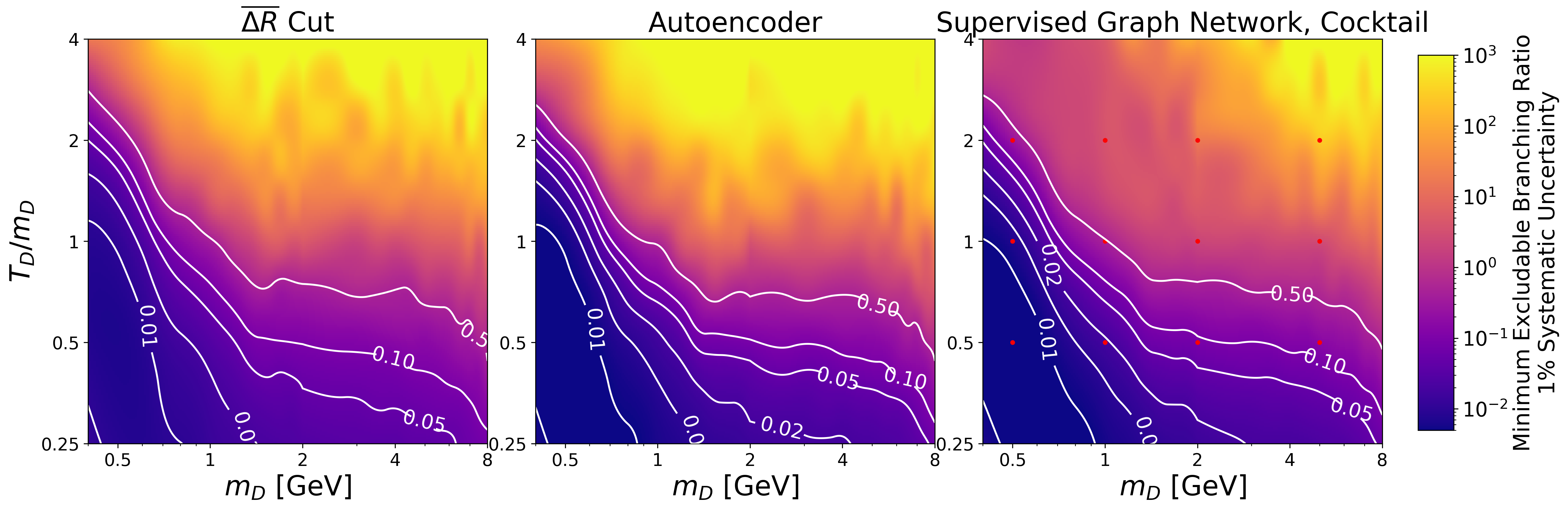}
    \\
    (a)\phantom{blablablablablablabla}(b)\phantom{blablablablablablabla}(c)\phantom{bla}
    \caption{
    Minimum excludable $\mathrm{Br}(h \to \mathrm{SUEP})$ at the HL-LHC, assuming $1\%$ systematic uncertainty on QCD background for (a) cut on $\overline{\Delta R}$, (b) fully connected autoencoder, (c) supervised graph network trained using cocktail of signal parameter choices (indicated with red dots). Note that the limited statistics of our QCD background sample leads these projections to  be very conservative while also de-emphasizing the performance differences between the three methods.
    }
    \label{fig:minbr}
\end{figure}

A key physics result is the actual sensitivity of HL-LHC searches to the SUEP final state. We therefore extract the smallest $\mathrm{Br}(h \to \ \mathrm{SUEP})$ branching ratio for which $S/\sqrt{B+u_{sys}B^{2}}>2$, where $u_{sys}$ gives the systematic uncertainty on the background.
Because of the high degree of background rejection required to be sensitive to relevant Higgs branching ratios, and the limited size of our background dataset, very few or no simulated background events remain when the classifier's threshold is set to maximize sensitivity to low branching ratios. This introduces a significant statistical uncertainty to the estimated LHC reach. We can decouple our estimate somewhat from these limitations by demanding the statistical uncertainty from the limited size of the background dataset to remain below $50\%$. This is conservative, since a realistic analysis will employ even harsher cuts with better significance improvement. 
For the same reason, the difference in reach between our three methods is likely underestimated. The performance gaps between the $\overline{\Delta R}$ cut, autoencoder, and supervised network are more clearly shown by the individual significance improvement curves and the AUC differences.

Finally, \fref{fig:minbr} shows the SUEP sensitivity achievable by the $\overline{\Delta R}$ cut, the autoencoder, and the cocktail-trained supervised network, all assuming 1\% systematic uncertainty on the background. Both the cut and the autoencoder can probe \%-level branching ratios.
In the Appendix we show sensitivity projections assuming a much larger 10\% systematic background uncertainty, with only very modest degradation in reach. This shows that the overwhelming QCD background has been reduced to low enough levels to make the search statistics limited, and speaks to the robustness of our results. 

%%%%%%%%%%%%%%%%%%%%%%%%%%%%%%
%%%%%%%%%%%%%%%%%%%%%%%%%%%%%%
%%%%%%%%%%%%%%%%%%%%%%%%%%%%%%
%%%%%%%%%%%%%%%%%%%%%%%%%%%%%%
%%%%%%%%%%%%%%%%%%%%%%%%%%%%%%
%%%%%%%%%%%%%%%%%%%%%%%%%%%%%%
%%%%%%%%%%%%%%%%%%%%%%%%%%%%%%
\section{Conclusion}\label{sec:Summary}

SUEPs represent a highly plausible but extremely challenging experimental signature of confining hidden sectors, which typically results in a high multiplicity of soft SM final states. To date, there are no targeted LHC searches for SUEP, and most existing searches have limited or no sensitivity. Furthermore, since SUEP is produced by hidden sectors featuring fairly strongly-coupled, approximately conformal dynamics with a wide variety of possible dark hadronization scenarios, modelling the detailed production of SUEP is fraught with uncertainties. Existing proposals for SUEP searches~\cite{Knapen:2016hky, Alimena:2019zri} target conspicuous, qualitative features of the final state, such as displaced vertices or leptons. Their existence is a  prediction for some decay portals of the dark hadrons, and targeting them with searches is fairly robust with respect to modelling uncertainties. 

An important but previously unaddressed question is how well we can look for SUEP using only its basic kinematic features, without any conspicuous SM final states.
This is not only an experimental challenge, the design of such a search also has to be very mindful of uncertainties in the signal modelling.
We investigated this worst-case scenario by studying SUEPs from dark hadrons produced in exotic Higgs decays, which decay promptly and purely hadronically. 
Our choice of model is motivated by the Higgs portal being one of the most plausible production modes for new physics. The modest energy scale of the Higgs decay also eliminates high-energy or high-mass observables as discriminants. Finally, Higgs production in association with a $W/Z$ circumvents the problem of triggering on the SUEP final state directly~\cite{Knapen:2016hky}.

Our first results are observables which capture the essential SUEP features from the track momenta. 
We focused on the charged particle multiplicity $N_\text{charged}$, the event ring-isotropy~\cite{Cesarotti:2020hwb} 
$\mathcal{I}$, and the interparticle distance $\Delta R_{ij}$. All three are robust with respect to modelling uncertainties of the SUEP final state, since they capture the essential model features: high multiplicity of final states, and dark hadron production that is more isotropic and democratic in momentum than the QCD background. The charged track interparticle distance matrix $\Delta R_{ij}$ is particularly suitable for ML applications. 

Based on these observables, we devised three strategies for $h \to \ \mathrm{SUEP}$ searches. The first is based on a simple cut on the interparticle distance matrix. The second assumes that our naive signal simulation tools can be trusted, and uses supervised ML techniques. The third is an unsupervised ML approach using a fully connected autoencoder trained only on simulated QCD events as an anomaly detector.
Both ML approaches used the slightly modified interparticle distance matrix of \eref{e.inputfeatures} as the event representation.
The cut-and-count approach serves as a simple baseline, over which the unsupervised ML approach represents a significant improvement, even without  detailed knowledge of the signal. This is to be compared to the higher sensitivity of the supervised machine learning approach, which is unlikely to be robust with respect to modeling uncertainties or choice of training parameters.

All three approaches will probe exotic Higgs decays to prompt, hadronic SUEP at the HL-LHC for branching fractions at the percent level, with both the autoencoder and supervised approaches probing rates as low as 1\%.
We assumed a systematic background uncertainty of one percent, but increasing this to ten percent only modestly decreases sensitivity, signaling that our analyses have sufficient differentiation power to reduce the enormous QCD background to a statistics-dominated level.  
These estimates are conservative, since our simulated background samples were still too small to fully cover the exceedingly large background rejection.
A realistic search combining the autoencoder with data-driven background estimates will achieve significantly higher sensitivities.

Our results show that even without a detailed theoretical description of the SUEP showering process, an analysis using an unsupervised neural network can be highly sensitive to exotic Higgs decays to SUEP. 
Our observables and techniques can equally well be used in SUEP searches using leptons, photons or displaced vertices, to significantly enhance their sensitivity based on the inherently SUEP-y kinematics.
Generalizations of our methods, for instance including searches for explicit dark hadron resonances, should allow for sensitivity to SUEPs with higher dark hadron masses or dark Hagedorn temperatures than those targeted by our analysis.
Our new, unsupervised search strategy can be applied to a wide range of LHC scenarios to discover new physics, even if the true BSM model should differ from our exact theory expectations.

\subsubsection*{Acknowledgements}
The authors would like to thank Anja Butter, Simon Knapen, Jessie Shelton, and Jennifer Thompson for helpful conversations.
The research of JB and DC is supported in part by a Discovery Grant from
the Natural Sciences and Engineering Research Council
of Canada, and by the Canada Research Chair program.
JB also acknowledges funding from a Postgraduate Doctoral Scholarship (PGS D) provided by the Natural Sciences and Engineering Research Council of Canada. 
GK acknowledges the support of the Deutsche Forschungsgemeinschaft (DFG, German Re\-search Foundation) 
under Germany’s Excellence Strategy – EXC 2121  ``Quantum Universe" – 390833306. 
The research of TP is supported by the Deutsche Forschungsgemeinschaft (DFG, German Research Foundation) under grant 396021762 -- TRR~257 \textsl{Particle Physics Phenomenology after the Higgs Discovery}. 

Computations were performed on the Niagara supercomputer at the SciNet HPC Consortium \cite{Niagara,SciNetLessons}. SciNet is funded by: the Canada Foundation for Innovation; the Government of Ontario; Ontario Research Fund - Research Excellence; and the University of Toronto. This research was enabled in part by support provided by Compute Canada (www.computecanada.ca).

\appendix

\section{Additional Results}

\begin{figure}[b!]
	\centering
	\includegraphics[width=\textwidth]{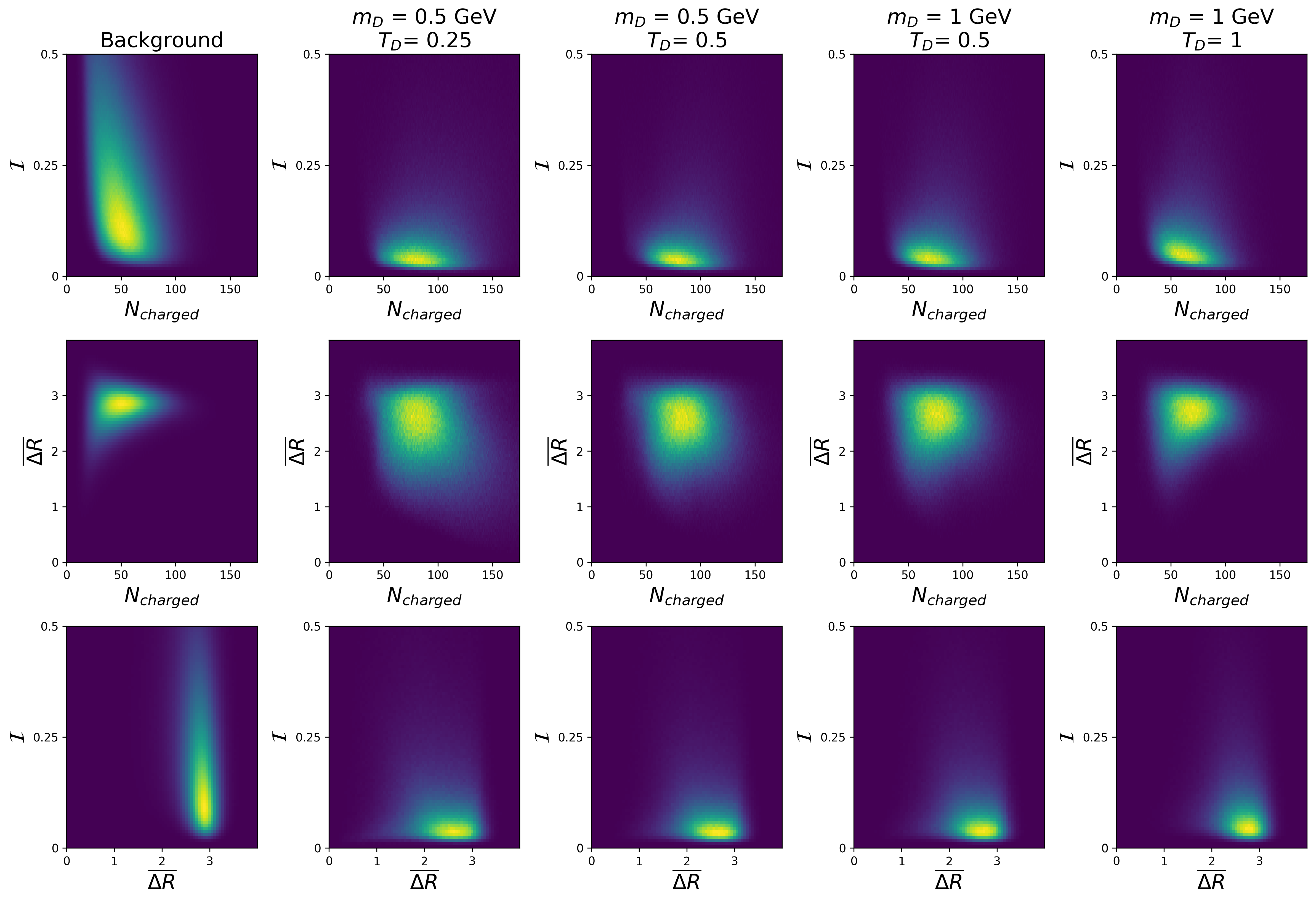}
	\hfill
	\caption{Two-dimensional unit-normalized histograms showing pairwise correlations between $N_{charged}$, $\mathcal{I}$, and $\overline{\Delta R}$ for the background sample and the same four choices of signal parameters as in Figure \ref{fig:ROC_comparison}.}
	\label{fig:observables_correlation}
\end{figure}

\begin{figure}
    \centering
    \includegraphics[width=\textwidth]{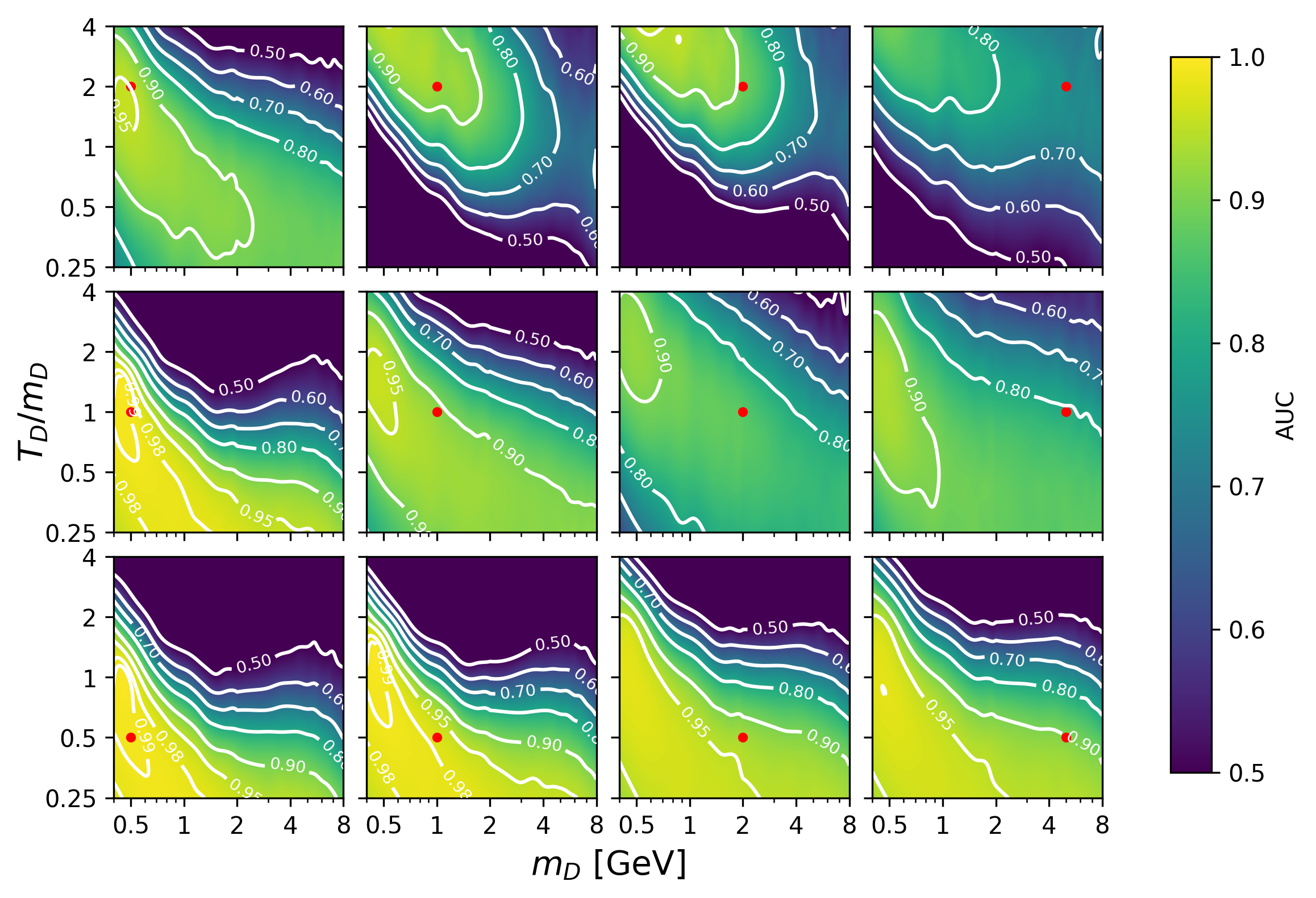}
    \hfill
    \caption{AUC for supervised graph networks trained using different signal parameter choices. Red dots indicate the training parameters in each plot.}
    \label{fig:supervised_auc}
\end{figure}

\begin{figure}
    \centering
    \includegraphics[width=\textwidth]{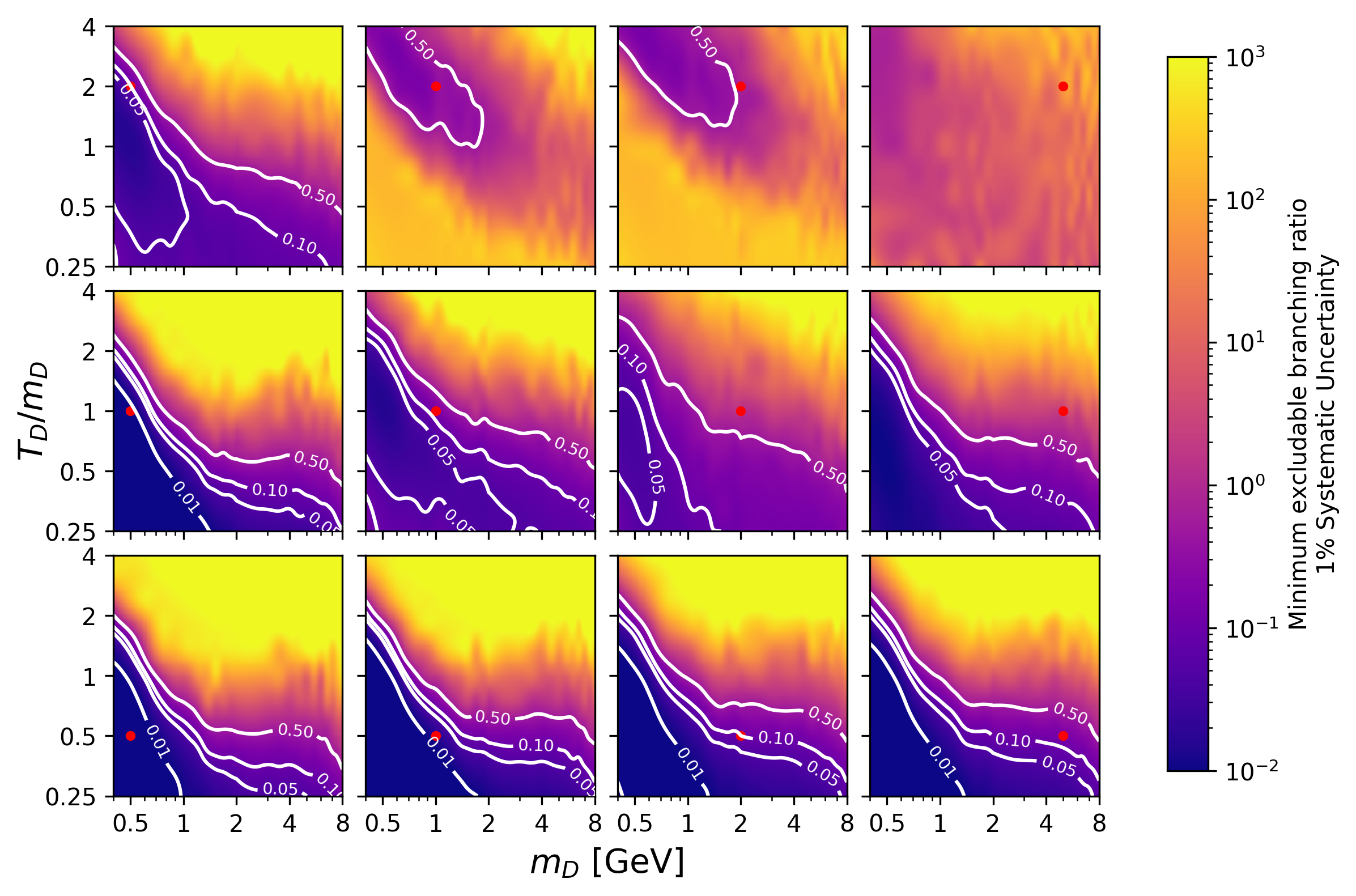}
    \caption{Minimum branching ratio excludable by supervised graph networks trained on different choices of ($m_{D}$,$T_{D}$). Red dots indicate the training parameters in each plot.}
    \label{fig:supervised_minbr}
\end{figure}

Figure \ref{fig:observables_correlation} shows the two-dimensional correlations between each pair of the observables $N_{charged}$, $\mathcal{I}$, and $\overline{\Delta R}$. While the average values of $N_{charged}$ and $\overline{\Delta R}$ vary in a similar fashion with $m_{D}$ and $T_{D}$, their distributions for a fixed choice of signal parameters are not highly correlated. 

Figure \ref{fig:supervised_auc} shows the AUC for each of the twelve supervised graph networks trained on a single sample of signal events, with parameters indicated by the red dot in each plot. This vividly demonstrates how the the parameter choice of the signal dataset on which the supervised network is trained has a large impact on the regime of parameter space where it can achieve high AUC, providing a strong argument for using the cocktail approach. 

The sensitivity projection for these supervised networks with one percent background systematic is shown in \fref{fig:supervised_minbr}. The peak sensitivity for each network is also at the 1\%-level, similar to the other approaches, but for potentially a different range of SUEP parameters, depending greatly on the training parameters of each network. 

\begin{figure}
    \centering
    \centering
    \includegraphics[width=0.9\textwidth]{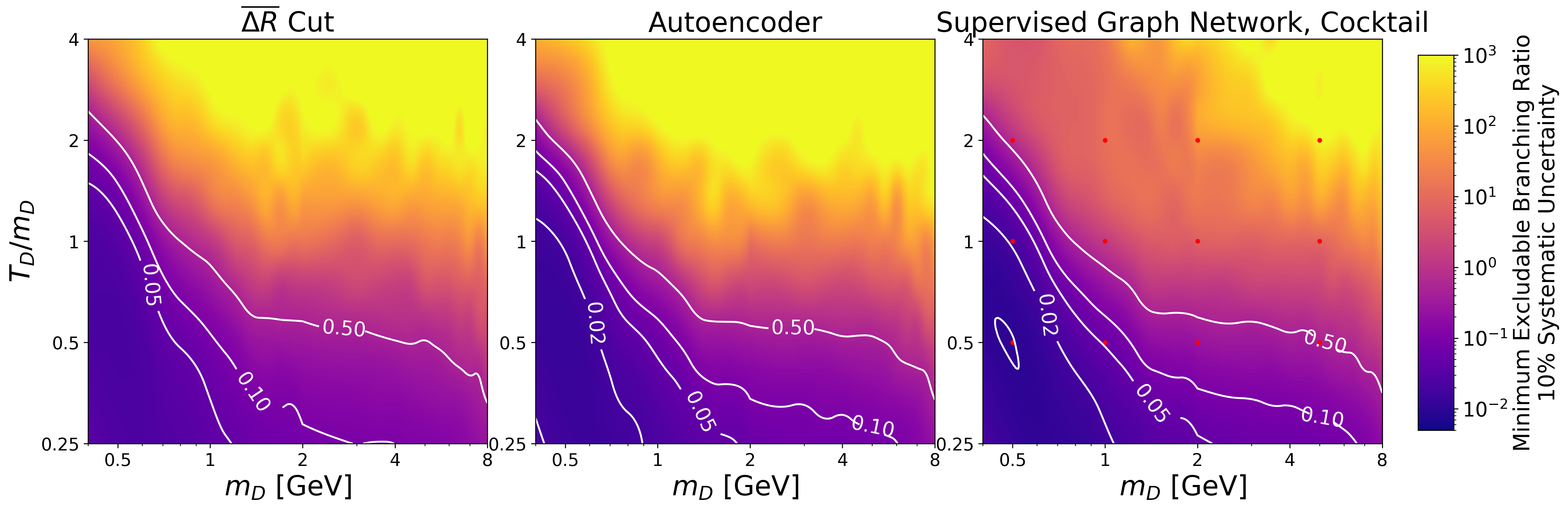}
    \\
    (a)\phantom{blablablablablablabla}(b)\phantom{blablablablablablabla}(c)\phantom{bla}
    \caption{
    Same as Fig.~\ref{fig:minbr} but for 10\% systematic background uncertainty.}
    \label{fig:minbr10}
\end{figure}

\begin{figure}
    \centering
    \includegraphics[width=\textwidth]{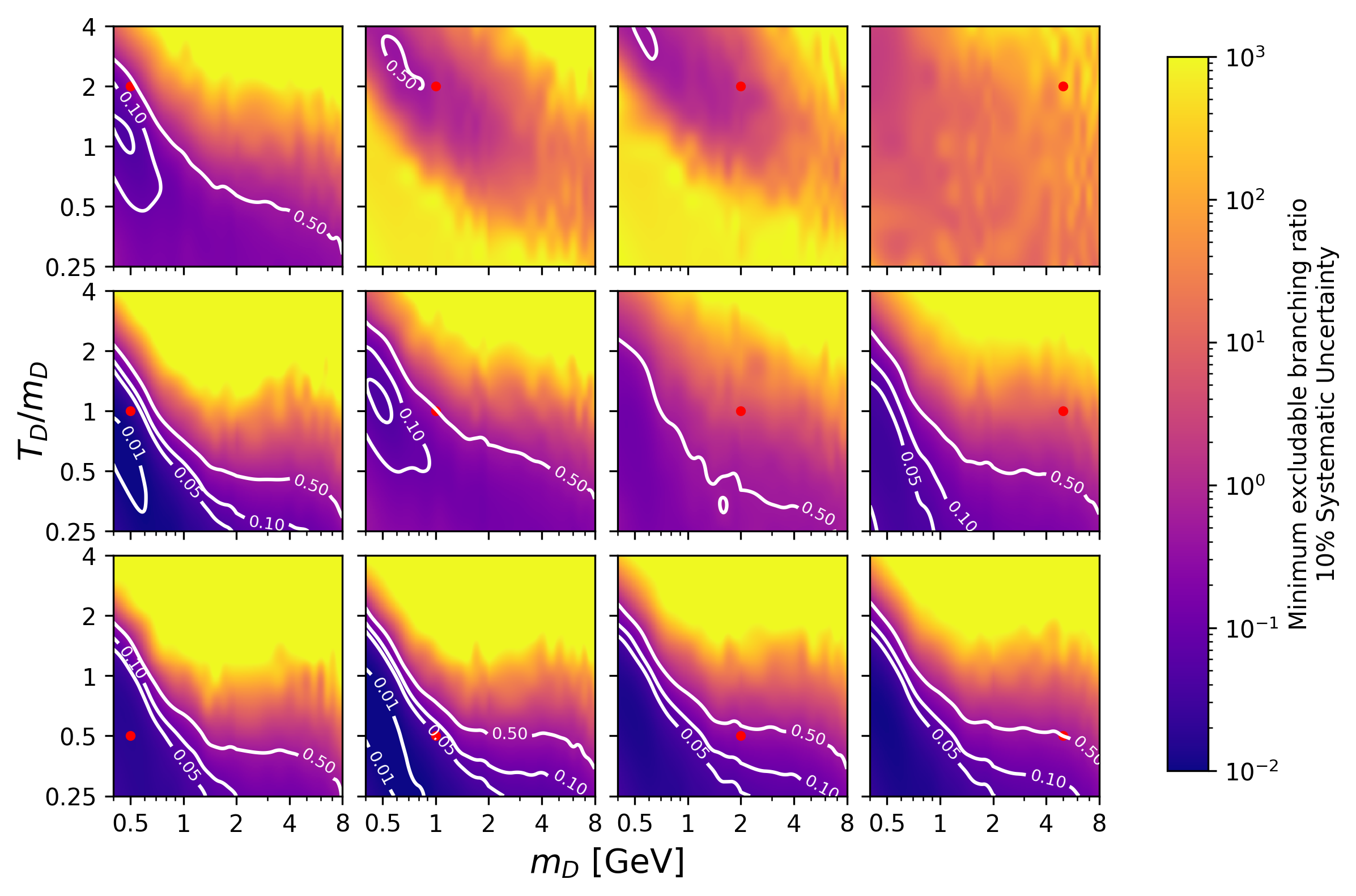}
    \caption{Same as Fig.~\ref{fig:supervised_minbr} but for 10\% systematic background uncertainty. }
    \label{fig:supervised_minbr10}
\end{figure}

Figures~\ref{fig:minbr10} and~\ref{fig:supervised_minbr10} show sensitivity projections for 10\% systematic background uncertainty.

\clearpage

\bibliographystyle{Jhep}
\bibliography{references}

\end{document}